\documentclass[twocolumn]{aastex631}

\usepackage{comment}
\usepackage{mathrsfs}
\usepackage[utf8]{inputenc}
\usepackage{newunicodechar}
\DeclareRobustCommand{\okina}{%
  \raisebox{\dimexpr\fontcharht\font`A-\height}{%
    \scalebox{0.8}{`}%
  }%
}
\newunicodechar{ʻ}{\okina}

\usepackage{natbib}
\usepackage{graphicx}
\usepackage{bigints}
\graphicspath{{./Figure}}

\usepackage{url}
\usepackage{hyperref} 
\usepackage{xcolor}
\usepackage{textcomp}

\usepackage{amsmath,amssymb}

\def\units#1{\hbox{ $\,{\rm #1}$}}
\def\degrees{\hbox{${}^\circ$} }

\usepackage{wasysym}

\begin{document}

\title{Constraints on the gamma-ray emission from Small Solar System Bodies with the Fermi Large Area Telescope data}

\author[0000-0002-3358-2559]{S.~De~Gaetano}
\affiliation{Istituto Nazionale di Fisica Nucleare, Sezione di Bari, via Orabona 4, I-70126 Bari, Italy}
\affiliation{Dipartimento di Fisica dell'Universit\`a e del Politecnico di Bari, via Amendola 173, I-70126 Bari, Italy}

\author[0000-0003-0703-824X]{L.~Di~Venere}
\affiliation{Istituto Nazionale di Fisica Nucleare, Sezione di Bari, via Orabona 4, I-70126 Bari, Italy}

\author[0000-0002-5055-6395]{F.~Gargano}
\affiliation{Istituto Nazionale di Fisica Nucleare, Sezione di Bari, via Orabona 4, I-70126 Bari, Italy}

\author[0000-0002-1173-5673]{F. Loparco}
\affiliation{Istituto Nazionale di Fisica Nucleare, Sezione di Bari, via Orabona 4, I-70126 Bari, Italy}
\affiliation{Dipartimento di Fisica dell'Universit\`a e del Politecnico di Bari, via Amendola 173, I-70126 Bari, Italy}

\author[0000-0002-2549-4401]{L.~Lorusso}
\affiliation{Istituto Nazionale di Fisica Nucleare, Sezione di Bari, via Orabona 4, I-70126 Bari, Italy}
\affiliation{Dipartimento di Fisica dell'Universit\`a e del Politecnico di Bari, via Amendola 173, I-70126 Bari, Italy}

\author[0000-0001-9325-4672]{M. N. Mazziotta}
\affiliation{Istituto Nazionale di Fisica Nucleare, Sezione di Bari, via Orabona 4, I-70126 Bari, Italy}

\author[0000-0002-2586-1021]{G.~Panzarini}
\affiliation{Istituto Nazionale di Fisica Nucleare, Sezione di Bari, via Orabona 4, I-70126 Bari, Italy}
\affiliation{Dipartimento di Fisica dell'Universit\`a e del Politecnico di Bari, via Amendola 173, I-70126 Bari, Italy}

\author[0000-0003-3808-963X]{R.~Pillera}
\affiliation{Istituto Nazionale di Fisica Nucleare, Sezione di Bari, via Orabona 4, I-70126 Bari, Italy}
\affiliation{Dipartimento di Fisica dell'Universit\`a e del Politecnico di Bari, via Amendola 173, I-70126 Bari, Italy}

\author[0000-0002-9754-6530]{D.~Serini}
\affiliation{Istituto Nazionale di Fisica Nucleare, Sezione di Bari, via Orabona 4, I-70126 Bari, Italy}

\correspondingauthor{S. De Gaetano, L. Di Venere and M. N. Mazziotta}
\email{salvatore.degaetano@ba.infn.it, leonardo.divenere@ba.infn.it, mazziotta@ba.infn.it}

\begin{abstract}
All known Small Solar System Bodies have diameters between a few meters and a few thousands of kilometers. Based on the collisional evolution of Solar System Bodies, a larger number of asteroids with diameters down to $\sim 2 \units{m}$ is thought to exist. As all Solar System Bodies, Small Bodies can be passive sources of high-energy gamma rays, produced by the interaction of energetic cosmic rays impinging on their surfaces. Since the majority of known asteroids are in orbits between Mars and Jupiter (in a region known as the Main Belt), we expect them to produce a diffuse emission close to the ecliptic plane. In this work we have studied the gamma-ray emission coming from the ecliptic using the data collected by the Large Area Telescope onboard the Fermi satellite. We have fit the results with simulations of the gamma-ray intensity at source level (calculated with the software {\tt FLUKA}) to constrain the Small Solar System Bodies population. Finally, we have proposed a model describing the distribution of asteroid sizes and we have used the LAT data to constrain the gamma-ray emission expected from this model and, in turn, on the model itself.
\end{abstract}

\section{Introduction}
\label{sec:intro}

The Small Solar System Bodies (SSSBs) include asteroids, comets, small planetary satellites and all the other objects in the Solar System which are not planets, dwarf planets or natural satellites. These bodies can be mainly divided into three families: the Main Belt, including all small bodies lying between the orbits of Mars and Jupiter; the Trojans, which share an orbit with a larger planet or moon; and the Kuiper Belt, made of Trans-Neptunian objects. 
Asteroids are classified based on their color, albedo and spectral types~\citep{lodders}.  About $75\%$ of known asteroids belong to the C-type class. These asteroids are extremely dark, since their composition includes carbon in addition to rocks and minerals. The second most abundant taxonomic species are S-type asteroids, which represent $17\%$ of the whole asteroids population. These asteroids are moderately bright and consist mainly of iron and magnesium silicates.
Finally, most of the remaining asteroids belong to the M-type class and are rich in metals (mainly iron and nickel).

All known asteroids have diameters $> 2 \units{m}$ (assuming a spherical shape) and the majority is distributed along the ecliptic plane. As for all other objects in the Solar System, these bodies can be passive sources of gamma rays, produced by inelastic interactions of cosmic rays impinging on them. The result is the production of a diffuse gamma-ray emission along the ecliptic plane which, if observed, could provide a way to further investigate the asteroid properties and, in particular, to study the distribution of their sizes.

In the present work, we have studied the gamma-ray flux from the ecliptic plane using the data collected by the Fermi Large Area Telescope (LAT) from August 2008 to December 2020 and we have used these results to constrain the gamma-ray emission from SSSBs. In addition, we have used the {\tt FLUKA} code to predict the gamma-ray emission resulting from CRs interacting with different types of asteroids. We have then used the analysis results to constrain the total number of asteroids with given properties. Finally, we have fit the LAT data with a diffuse flux model of the SSSBs, obtained by folding the gamma-ray intensity calculated with {\tt FLUKA} with a population distribution function obtained by extending a model proposed by ~\citet{davis2002collisional} to diameters down to $\simeq 20 \units{cm}$. We have used the fit results to constrain the parameters of the diffuse flux and, in turn, the above-mentioned population model.

The idea of probing asteroid populations using gamma-ray observations was already considered by~\citet{Moskalenko:2007tk} and~\citet{Moskalenko:2009tv}, who also calculated the expected gamma-ray fluxes from asteroids under some simplifying assumptions. In the present work we propose a model that extends the previous ones, by including a description of the spatial morphology of the gamma-ray emission from asteroids.

\section{Small Solar System Bodies}
\label{sec:smallsolarsystembodies}

The asteroid mass and size distributions are thought to result from collisions during their evolution and accretion. Collisions between asteroids give rise to a cascade of fragments, shifting masses toward smaller sizes, while slow accretion leads to the asteroid growth~\citep{Dohnanyi1969}. Under these assumptions, the size distribution can be described with a power-law model:

\begin{equation}
    \frac{dN}{dr} = a ~ r^{-\alpha}
    \label{eq:dndr}
\end{equation}
where asteroids are modelled as spheres of radius $r$ and the power-law index is $\alpha \simeq 2.7$.

Assuming that all the asteroids are homogeneous bodies with the same density $\rho$ and a spherical shape, their mass distribution ($m=\frac{4 \pi \rho}{3} r^3$) is also described by a power-law:

\begin{equation}
    \cfrac{dN}{dm} = \cfrac{dN}{dr} \cfrac{dr}{dm} = \cfrac{a}{3} \left( \cfrac{4 \pi \rho}{3} \right)^{\frac{\alpha-1}{3}} m^{-\frac{\alpha+2}{3}}= b~ m^{-\kappa}
\end{equation}
with $\kappa = \frac{\alpha+2}{3}$ and $b=\frac{a}{3} \left( \frac{4 \pi \rho}{3} \right)^{\frac{\alpha-1}{3}}$.

The parameter $a$ can be calculated from the total mass $M$ of the whole asteroid population. In fact, assuming that asteroid masses are distributed in the range from $m_{0}$ to $m_{1}$, the total mass is given by:

\begin{equation}
    M = \int_{m_0}^{m_1} m \frac{dN}{dm} dm = 
    \begin{cases} 
    b \cfrac{m_1^{2-\kappa}-m_0^{2-\kappa}}{2-\kappa}  & \text{for $\kappa \neq 2$} \\ \\
    b \log \cfrac{m_1}{m_0} & \text{for $\kappa = 2$}
    \end{cases}
    \label{eq:tmass}
\end{equation}

The parameter $a$ is therefore given by:

\begin{equation} 
 a =  
 \begin{cases}
 \cfrac{3M}{4 \pi \rho} \times \cfrac{4-\alpha}{r_1^{4-\alpha}-r_0^{4-\alpha}}  & \text{for $\alpha \neq 4$} \\ \\
 \cfrac{3M}{4 \pi \rho} \times \cfrac{1}{ \log \frac{r_1}{r_0} } & \text{for $\alpha = 4$}
 \end{cases}
\label{eq:a}
\end{equation}
where $r_0$ and $r_1$ are respectively the radii of asteroids with mass $m_0$ and $m_1$.

The total mass of the asteroids in the Main Belt (semimajor axis $\simeq 2.7$ AU) and of Jovian Trojans (semimajor axis $\simeq 5.2$ AU) is estimated to be of about $10^{-4} - 10^{-3} M_\oplus$, where $M_\oplus$ is the mass of the Earth, while the total mass of the asteroids in the Kuiper Belt beyond Neptune (semimajor axis $\simeq 40-50$ AU) is estimated to be about  $10^{-2} M_\oplus$~\citep{pitjeva2018masses}~\footnote{The total asteroid mass is of the same order of magnitude as the mass of Moon, which is $\sim 10^{-2} M_\oplus$.}.

\begin{figure}[!t]
    \centering
    \includegraphics[width=0.95\columnwidth,clip]{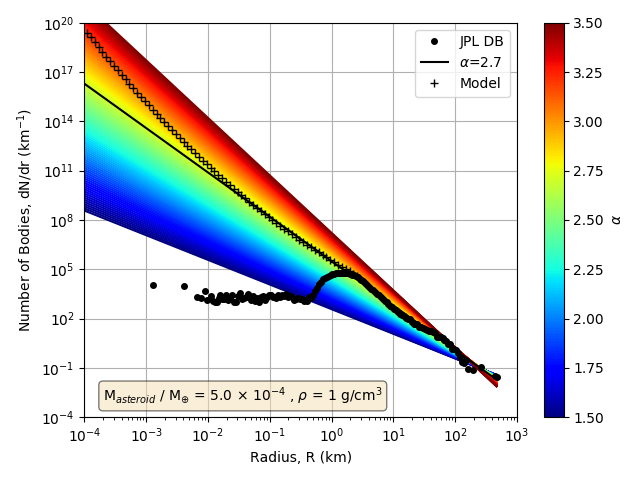}
    \caption{SSSBs size distribution. The distribution extracted from the JPL database is indicated with black circles, while the points of our model are indicated with black crosses. The colored bands show the power-law models (Eq.~\ref{eq:dndr}) for different values of $\alpha$, assuming that the total mass of asteroids is $5 \times 10^{-4}~M_\oplus$ and assuming an asteroid density of $1\units{g/cm^{3}}$. The black line shows the differential size distribution with $\alpha=2.7$.}
    \label{fig:ssbs_size}
\end{figure}

The NASA Jet Propulsion Laboratory (JPL) database catalog~\citep{JPL} provides the orbital parameters of almost $10^6$ small bodies along with their diameters. The Trojan population is numerically $10-25\%$ of the Main Belt population at sizes $\leq 100\units{km}$ and the smallest diameter reported in the catalog is about $10^{-3} \units{km}$. In Fig.~\ref{fig:ssbs_size} the differential distribution of the radii of SSSBs found in the JPL database is shown.  In~\citet{davis2002collisional} several estimates of the Main Belt asteroid size distribution down to smaller diameters are presented. 

The model in~\citet{durda1998} is a fit to the distribution determined by~\citet{JEDICKE1998245}, where the authors used the Spacewatch data to estimate the size distribution of SSSBs in the Main Belt. The estimate of the cumulative size distribution of asteroids (i.e., number of asteroids $N(d>D)$ with diameter $d$ greater than a certain value $D$) is given and $N(d>10^{-2}\units{km}) \simeq 10^{10}$.
In this work we extend the model down to diameters of $\simeq 20 \units{cm}$, extrapolating it with a log-parabola function. In particular, we assume the diameters to be distributed according to the JPL catalog for values above $2.5 \units{km}$, and to follow the extrapolated model of~\citet{durda1998} for diameters in the range $20 \units{cm} - 2.5 \units{km}$. The resulting model is shown in Fig.~\ref{fig:ssbs_size} as a differential size distribution. In the same figure, we also show the distributions obtained assuming a power-law model as in Eq.~\ref{eq:dndr} for different values of the parameter $\alpha$, with a total mass of the asteroids $M=5 \times 10^{-4} ~ M_\oplus$, $r_0 = 1 \times 10^{-4}$ km, $r_1 = 470 $ km (Ceres' radius) and assuming an asteroid density of $1\units{g/cm^3}$. In Fig.~\ref{fig:ssbs_size} the differential distribution of the radii of the bodies in the JPL catalog is also shown. The JPL catalog includes only observed objects; therefore the size distribution is underestimated for smaller radii as these objects are difficult to detect.

As explained in the following sections, the analysis of gamma rays detected by the LAT provides a way to set constraints on the population and size distribution of asteroids.

\section{Asteroids gamma-ray emission}
\label{sec:asteroidsspectrum}

As mentioned in the previous section, asteroids should produce a diffuse gamma-ray emission along the ecliptic plane due to interactions of charged cosmic rays with their surfaces. Hereafter we will assume the asteroids to be spherical. The gamma-ray flux produced by $N(r,d)$ asteroids of radius $r$ at distance $d$ from the Earth (in units of$\units{photons~GeV^{-1}~cm^{-2}~s^{-1}}$) is given by the following equation~\citep{Fermi-LAT:2016tkg,Mazziotta:2020uey}:

\begin{equation}
    \phi_\gamma(E_{\gamma},d,r)=\pi \frac{r^2}{d^2}I_{\gamma}(E_{\gamma},r) N(r,d)
\label{eq:astflux}
\end{equation}
where $E_{\gamma}$ is the gamma-ray energy and $I_{\gamma}$ is the differential intensity of gamma rays at the production site.

\subsection{Gamma-ray intensity at production}
\label{subsec:Igamma}

The gamma-ray intensity at production for an asteroid of radius $r$, in units of$\units{photons~GeV^{-1}~cm^{-2}~sr^{-1}~s^{-1}}$, is given by:

\begin{equation}
    I_{\gamma}(E_{\gamma},r) = \sum_i \int Y_i(E_{\gamma}|E_k, r) I_i(E_k) dE_k
\label{eq:astint}
\end{equation}
where $I_i(E_k)$ is the intensity of the i-th species of cosmic rays impinging on the asteroid surface (mostly protons, electrons and He nuclei) and $Y_i(E_{\gamma}|E_k, r)$ is the yield of gamma rays produced by the interaction of the $i$-th cosmic-ray species with kinetic energy $E_k$ with the body surface of radius $r$. 

We have calculated the yield $Y_i(E_{\gamma}|E_k, r)$ using the {\tt FLUKA} code~\citep{Ferrari:2005zk,BOHLEN2014211,BATTISTONI201510}. {\tt FLUKA} is a general purpose Monte Carlo code for the simulation of hadronic and electromagnetic interactions, used in many applications. It can simulate with high accuracy the interactions and propagation in matter of about 60 different species of particles, including photons and electrons from $1\units{keV}$ to thousands of$\units{TeV}$, neutrinos, muons of any energy, hadrons and the corresponding antiparticles of energies up to $20\units{TeV}$ or up to $10\units{PeV}$ when it is interfaced with the {\tt DPMJET} code~\citep{10.1007/978-3-642-18211-2_166}, neutrons down to thermal energies and heavy ions.

Hadronic interactions in {\tt FLUKA} below a few$\units{GeV}$ are based on resonance production and decay of particles, while for higher energies the Dual Parton Model is used, implying a treatment in terms of quark chain formation and hadronization. The interactions are simulated in the framework of the PreEquilibrium Approach to NUclear Thermalization model ({\tt PEANUT})~\citep{Fasso:2000hd,battistoni2006recent}, including the Gribov-Glauber multi-collision mechanism followed by the pre-equilibrium stage and eventually equilibrium processes (evaporation, fission, Fermi break-up and gamma deexcitation).
We refer the reader to~\citet{Mazziotta:2015uba} and references therein for a more extended description on the interaction models that {\tt FLUKA} employs for these interactions in different energy ranges. Full information on the different models used by the code and its related publications and references can be found in the FLUKA webpage~\footnote{\url{http://www.fluka.org/}.}.

The {\tt FLUKA} code already has been used to model the gamma-ray emission from the Moon~\citep{Fermi-LAT:2016tkg} and the Sun~\citep{Mazziotta:2020uey}, providing excellent agreement with data.

In our simulation setup, each SSSB is defined as a spherical body with radius ranging from $10\units{cm}$ to $100\units{km}$. 
We have simulated different kinds of bodies, with different compositions and densities.

We first defined three homogeneous bodies to investigate possible dependence on the simulated material and density:

\begin{itemize}
    \item Ice: H$_2$O with density of $0.92\units{g~cm^{-3}}$
    \item Silica: SiO$_2$ with density of $2.00\units{g~cm^{-3}}$
    \item Carbon: C with density of $2.00\units{g~cm^{-3}}$
\end{itemize}

For ice, we assumed the same density as on the Earth. For silica and carbon bodies we chose the same density, in order to better investigate possible effects specifically due to the element.

Secondly, we defined two species which are representative of the most abundant asteroids:

\begin{itemize}
    \item C-type asteroids, with a density of $2.23\units{g~cm^{-3}}$
    \item S-type asteroids, with a density of $3.80\units{g~cm^{-3}}$
\end{itemize}

We assumed that the C-type and S-type composition is the same as the one of carbonaceous and ordinary chondrites respectively, with the elemental abundances taken from tables 16.10 and 16.11 of~\citep{lodders}.

To evaluate the yields of secondary particles from the SSSBs we have simulated several samples of protons, electrons and $^{4}$He nuclei with different kinetic energies, impinging on the asteroids with an isotropic and uniform distribution. The primary kinetic energy values are taken on a grid of $81$ equally spaced values in a logarithmic scale, from $100\units{MeV/n}$ up to $10\units{TeV/n}$.

The differential yield of secondary particles produced by the $i$-th species of cosmic-ray primaries (here $i=p$, $e^-$ and $^{4}$He), $Y_{i}(E_{\gamma} | E_k, r)$, is calculated by counting the secondary particles which escape from the asteroid. The yield in units of GeV$^{-1}$ is defined as:

\begin{equation}
Y_{i}(E_{\gamma} |~ E_k, r) = \frac{N_{i}(E_{\gamma} |~ E_k, r)}
{N_{i}(E_k) \Delta E_{\gamma}}
\label{eq:yield}
\end{equation} 
where $N_{i}(E_k)$ is the number of primaries of the $i$-th species generated with kinetic energy $E_k$ ($E_k$ is expressed in units of$\units{GeV}$ for primary electrons and protons and of$\units{GeV/n}$ for primary nuclei) and $N_{i}(E_{\gamma}|~E_k, r)$ is the number of photons with energies between $E_{\gamma}$ and $E_{\gamma} + \Delta E_{\gamma}$ produced by the primaries of the type $i$ with kinetic energy $E_k$ and escaping from the asteroid.

\begin{figure*}[!t]
    \centering
    \includegraphics[width=0.95\columnwidth,clip]{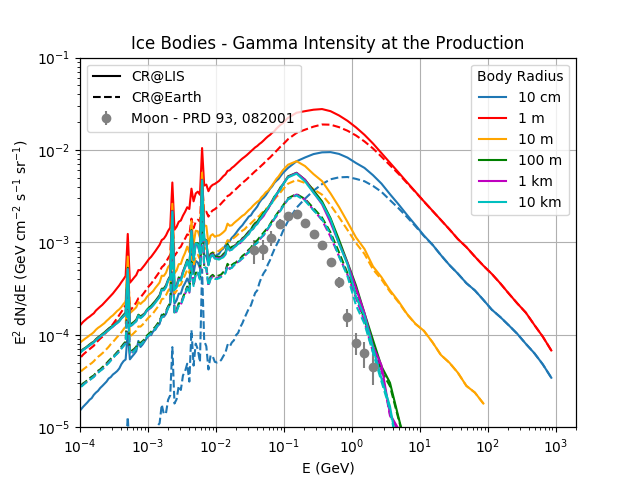}
    \includegraphics[width=0.95\columnwidth,clip]{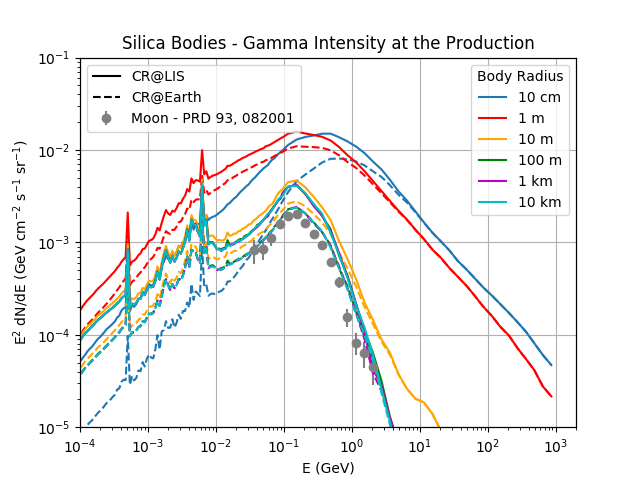}
    \includegraphics[width=0.95\columnwidth,clip]{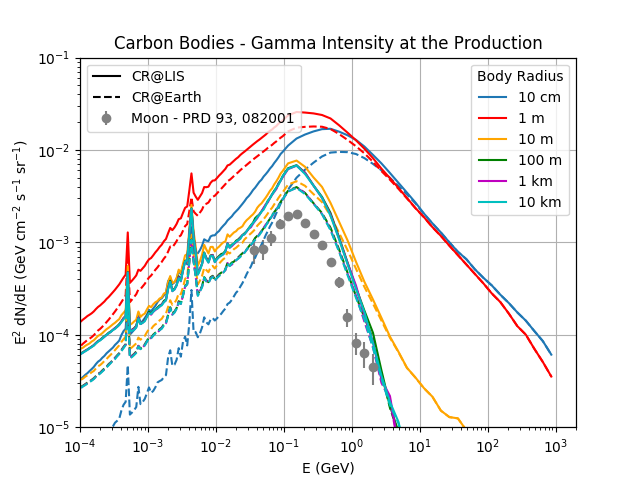}
    \includegraphics[width=0.95\columnwidth,clip]{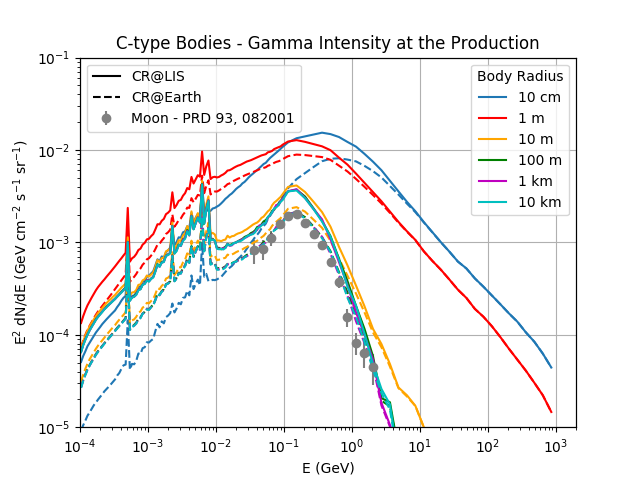}
    \includegraphics[width=0.95\columnwidth,clip]{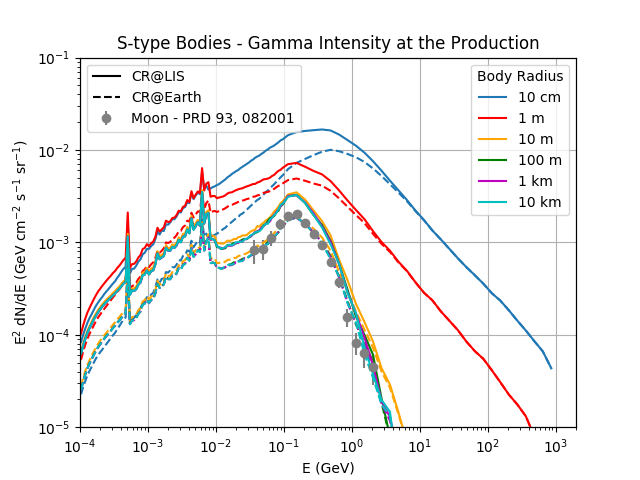}
    \caption{Gamma-ray intensities (multiplied by the energy squared) from ice, silica, carbon, C-type and S-type asteroids. The spectra have been evaluated assuming different radii, from $10\units{cm}$ to $10\units{km}$. The calculations have been performed in the two limiting cases, using the CR LIS spectra (continuous lines) and the CR spectra measured at Earth (dashed lines). The gray points show the intensity of the Moon measured by the Fermi-LAT~\citep{Fermi-LAT:2016tkg}.}
    \label{fig:inten}
\end{figure*}

The intensities $I_i(E_k)$ at the asteroid position can be calculated starting from the local interstellar spectra (LIS) taking into account the propagation of CRs in the Solar System, which is affected by solar activity. The activity of the Sun modulates the CR spectra with a 11-year cycle and its effect depends on the position of the asteroid in the Solar System. In this work we assume two limiting classes of CR spectra: the first one is given by the LIS, i.e. the spectra which are not affected by solar modulation, while the second are those measured at the Earth, where the solar modulation effect is larger than that at the positions of all asteroids, since their orbits are external to the Earth.
In this way we are bracketing the asteroid emission between these two limiting cases, since asteroids are distributed at distances of 50 AU and beyond (see Fig.~\ref{fig:orbitdist}). We have taken the CR LIS from~\citet{DeLaTorreLuque:2021yfq,Luque:2021nxb,Luque:2022aio}. The CR spectra at the Earth have been precisely measured by the AMS-02 instrument. In particular, we have taken the proton spectrum from~\citet{Aguilar:2015ooa}, the helium spectrum from~\citet{Aguilar:2015ctt,Aguilar:2017hno} and the electron+positron spectrum from~\citet{Aguilar:2014fea}. For the electron+positron spectrum at high energies we have also used the measurements by the Fermi-LAT~\citep{Abdollahi:2017nat} and DAMPE~\citep{Ambrosi:2017wek}~\citep[see also ][]{Mazziotta:2020uey}~\footnote{We did not simulate electrons and positrons separately, but only primary electrons assuming a spectrum equal to the overall electron+positron one.}.

In addition, we expect a time-dependent gamma-ray signal due to the 11-year solar cycle, which modulates the cosmic-ray intensities. In particular, this modulation is observed in the Moon gamma-ray flux, with variations over one Solar cycle in the range $\pm$15\% of the average emission~\citep{Fermi-LAT:2016tkg,DeGaetano:2021IG}.

Fig.~\ref{fig:inten} shows the gamma-ray intensities (Eq.~\ref{eq:astint}) evaluated for the different classes of asteroids simulated in this work~\footnote{As an example, in appendix~\ref{sec:appendix} we report detailed plots of gamma-ray yields and intensities for the different particle species interacting with Silica asteroids.}. Most lines in the spectra are due to photons emitted in nuclear de-excitation processes. The line at $511\units{keV}$ is due to annihilations of positrons produced in the electromagnetic showers.
The line at $2.2\units{MeV}$, which is visible in the ice bodies, is due to neutron capture by hydrogen nuclei, with the production of a deuterium nucleus and the emission of a gamma ray. Other classes of asteroids exhibit characteristic lines related to their composition. We remark that these features can be of particular interest for in-situ studies of asteroid composition. The gamma-ray intensities from the different classes of asteroids are also compared with the gamma-ray intensity from the Moon measured by the Fermi LAT~\citep{Fermi-LAT:2016tkg}. We see that the intensity of gamma rays emitted from silica bodies of large sizes is close to the intensity of gamma rays emitted from the Moon.

\begin{figure}[t]
    \centering
    \includegraphics[width=0.9\columnwidth,clip]{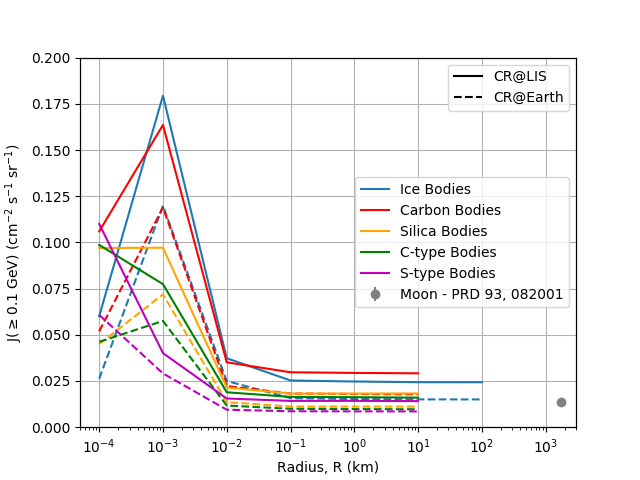}
    \caption{Gamma-ray intensity above $0.1\units{GeV}$ as a function of the body radius using the CR LIS spectra (continuous lines) and the CR spectra measured at Earth (dashed lines). The colored lines indicate the different compositions. The gray point shows the Moon data measured by the Fermi-LAT~\citep{Fermi-LAT:2016tkg}.}
    \label{fig:j100}
\end{figure}

Fig.~\ref{fig:j100} shows the integral of the intensity above $0.1\units{GeV}$ as a function of the asteroid radius for the different classes of asteroids. It can be noticed that the intensity at the production site drops for radii smaller than $1 \units{m}$, since the asteroid size becomes comparable or smaller than the typical interaction length in the simulated materials, which are of the order of tens of centimeters. From Figs.~\ref{fig:inten} and~\ref{fig:j100} we see that for $r>10 \units{m}$ the shapes of the spectra (and consequently the integral of the intensity above $0.1\units{GeV}$) do not depend on the asteroid radius and are similar to the shape of the gamma-ray intensity from the Moon~\citep{Fermi-LAT:2016tkg}. This is because the secondary gamma rays produced by cosmic rays impinging on the asteroids can escape only from external layers, since the cosmic-ray nuclei can penetrate down to depths of a few tens of grams per centimeter squared (corresponding to the hadronic interaction length for protons and He nuclei), while the gamma-ray absorption length (corresponding to the radiation length) is shorter. When the asteroid size is larger than both of these characteristic lengths, the gamma-ray production becomes independent of the size. 

\subsection{Gamma-ray flux at the Earth}

\begin{figure}[t]
    \centering
    \includegraphics[width=0.95\columnwidth,clip]{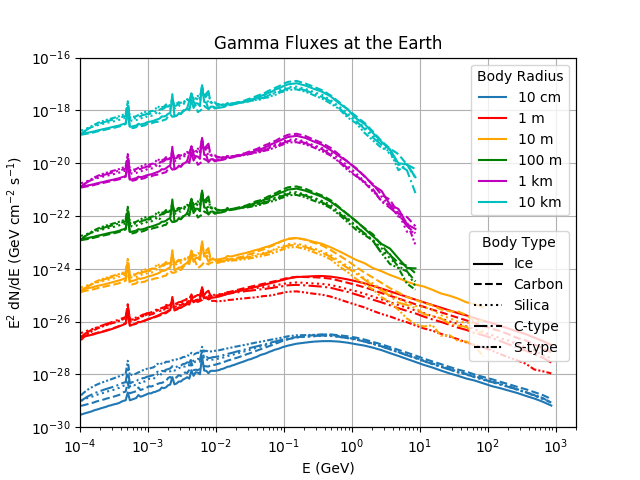}
    \caption{Gamma-ray fluxes at the Earth (multiplied by the energy squared) from single spherical bodies of different radii and types at the distance of $2.7 \units{AU}$ from the Earth.}
    \label{fig:ssbs_fluxspectrum}
\end{figure}

Fig.~\ref{fig:ssbs_fluxspectrum} shows the gamma-ray fluxes at the Earth from asteroids of different classes and different radii at a distance of $2.7 \units{AU}$ from our planet, evaluated using Eq.~\ref{eq:astflux} with $N(r,d)=1$. 

The gamma-ray flux produced by asteroids of radius $r>r_0$
(cumulative flux) at a given distance $d$ is given by:

\begin{equation}
    \phi_\gamma(E_{\gamma},d,r>r_0)= 
    \displaystyle\int_{r_0}^{r_1} \frac{\pi r^2}{d^2} I_{\gamma}(E_{\gamma},r)  \frac{dN}{dr} dr ,
\label{eq:astcumflux}
\end{equation}
where $r_1 \sim 500 \units{km}$ is the largest radius of observed asteroids.

\begin{figure*}[t]
    \centering
    \includegraphics[width=0.95\columnwidth,clip]{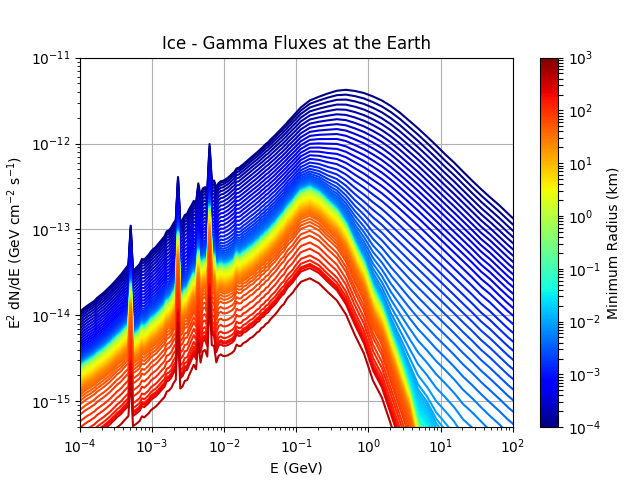}    
    \includegraphics[width=0.95\columnwidth,clip]{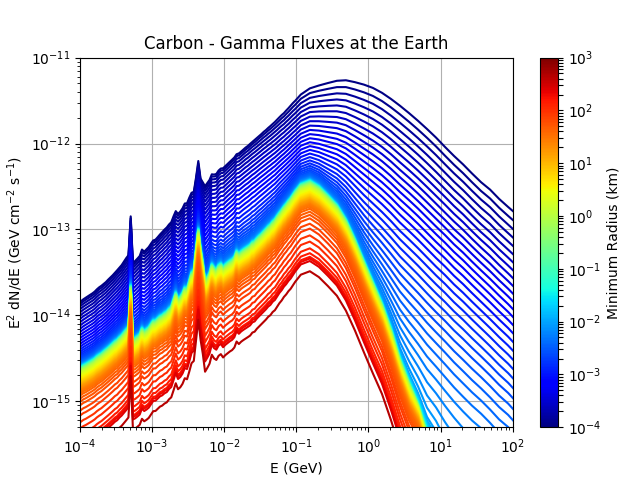}
    \includegraphics[width=0.95\columnwidth,clip]{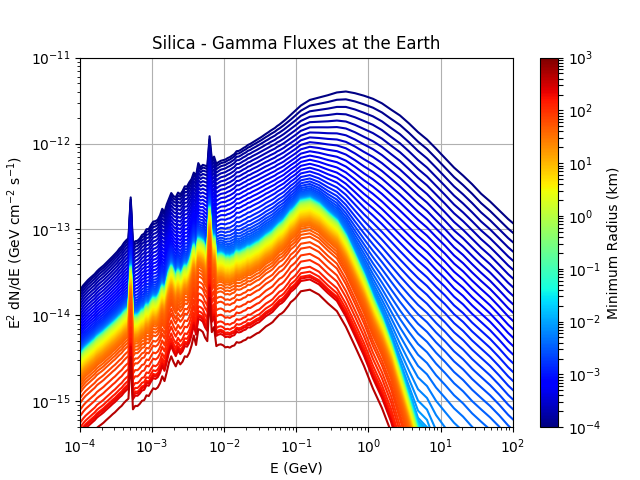}
    \includegraphics[width=0.95\columnwidth,clip]{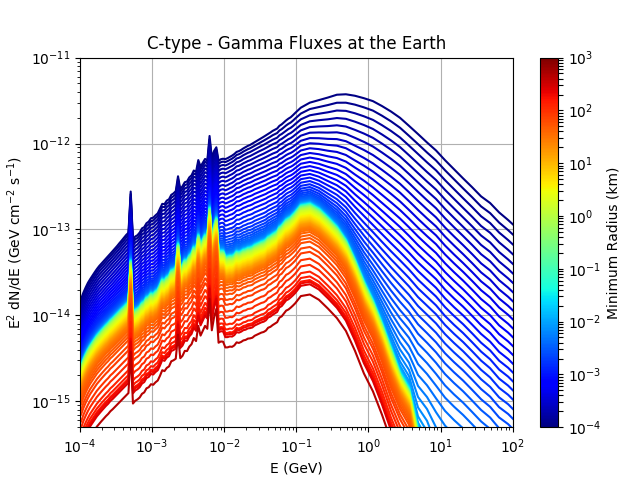}
    \includegraphics[width=0.95\columnwidth,clip]{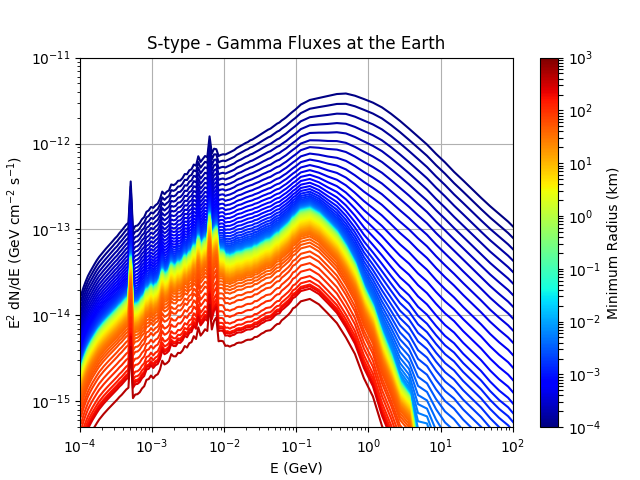}
    \caption{Cumulative gamma-ray fluxes (multiplied by $E_{\gamma}^{2}$) at the Earth for the SSSB population model described in section~\ref{sec:smallsolarsystembodies} and shown in Fig.~\ref{fig:ssbs_size}. Top-left panel: ice bodies; top-right panel: carbon bodies; middle-left panel: silica bodies; middle-right panel: C-type bodies; bottom panel: S-type bodies. The colored lines correspond to different values of the minimum asteroid radius $\bar{r}$.
    }
    \label{fig:ssbs_spec_ear}
\end{figure*}

Fig.~\ref{fig:ssbs_spec_ear} shows the cumulative gamma-ray fluxes at the Earth calculated with a population of SSSBs shown in Fig.~\ref{fig:ssbs_size} at a distance of $2.7\units{AU}$, for different values of $r_0$ in the range from $10\units{cm}$ to $10^{3} \units{km}$.

As discussed above, the gamma-ray emission at the site of production is almost independent of size for asteroid radii larger than a few tens of meters. Assuming that the gamma-ray intensity $I_{\gamma}$ does not depend on the asteroid radius $r$, the gamma-ray flux at the Earth for a power-law size distribution of the asteroids at a given distance $d$ can be expressed as:

\begin{equation}
\begin{array}{ll}
 \phi_{\gamma} & =   
 \pi I_{\gamma} \displaystyle\int_{r_0}^{r_1} \cfrac{r^2}{d^2}~ a r^{-\alpha} dr = \\ \\
 & = \begin{cases}
 \pi I_{\gamma} \cfrac{a}{d^2} \cfrac{r_1^{3-\alpha}-r_0^{3-\alpha}}{3-\alpha}  & \text{for $\alpha \neq 3$} \\ \\
 \pi I_{\gamma} \cfrac{a}{d^2} \log \cfrac{r_1}{r_0} & \text{for $\alpha = 3$}
\end{cases}
\end{array}
\label{eq:astflux1}
\end{equation}
where $r_{0}$ and $r_{1}$ are the minimum and maximum asteroid radii and the parameter $a$ can be calculated from Eq.~\ref{eq:a}. 

Following~\cite{Fermi-LAT:2016tkg}, the gamma-ray flux from the Moon can be expressed as:

\begin{equation}
    \phi_{\leftmoon} = \pi I_{\gamma} \cfrac{R_{\leftmoon}^{2}}{D_{\leftmoon}^{2}}
\label{eq:moonflux}
\end{equation}
where $I_{\gamma}$ is the intensity of lunar gamma rays, while $R_{\leftmoon}$ and $D_{\leftmoon}$ indicate the lunar radius and the Earth-Moon distance, respectively.

From the results shown in Figs.~\ref{fig:inten} and~\ref{fig:j100}, assuming that the gamma-ray intensity from asteroids is the same as from the Moon~\footnote{We note that the CR intensities are not the same at the asteroid and Moon positions, and therefore the gamma-ray intensities are slightly different (see discussion in Sect.~\ref{sec:asteroidsspectrum}).}, from eqs.~\ref{eq:astflux1} and~\ref{eq:moonflux} it follows that:

\begin{equation}
 \frac{\phi}{\phi_{\leftmoon}} = 
 \begin{cases}
 a \cfrac{1}{R_{\leftmoon}^2}
 \cfrac{D_{\leftmoon}^2}{d^2}  \times
 \cfrac{r_1^{3-\alpha}-r_0^{3-\alpha}}{3-\alpha}  & \text{for $\alpha \neq 3$} \\ \\
 a \cfrac{1}{R_{\leftmoon}^2}
 \cfrac{D_{\leftmoon}^2}{d^2}  \times
 \log \cfrac{r_1}{r_0} & \text{for $\alpha = 3$}
\end{cases}
\label{eq:ratioastmoonflux}
\end{equation}

We remark that Eq.~\ref{eq:ratioastmoonflux} differs from Eq. 6 in~\citet{Moskalenko:2007tk}, where it was assumed that the emission from small bodies scales with the radius of the body. In this work we calculate the gamma-ray flux from each body as in Eq.~\ref{eq:astflux}. For radii larger than $\sim 10 \units{m}$, the intensity at production becomes independent of the radius, as discussed above, and the flux at the Earth scales with the square of the radius. As a consequence, 
assuming a population of asteroids with radii between $r_0 = 1 \times 10^{-4} \units{km}$ and $r_1 = 470  \units{km}$, following Eq.~\ref{eq:ratioastmoonflux} our model predicts an asteroid flux from two to six orders of magnitudes lower than the flux calculated in Eq. 6 in~\citet{Moskalenko:2007tk}, depending on the index $\alpha$ of the power-law describing the asteroid size distribution.

\subsection{Spatial map of small bodies}
\label{subsec:spatialmap}

Since the orbits of the asteroids lie in an extended region of the sky, to build a template describing their gamma-ray emission we have first divided the sky into equal solid angle pixels and then we have added together the contributions from individual pixels. 

The differential gamma-ray flux from the SSSBs of radius $r$ in a sky pixel at the ecliptic coordinates $(\lambda,\beta)$, covering a solid angle $\Delta \Omega$ is given by:

\begin{equation}
    \phi_\gamma(E_{\gamma},r,\lambda,\beta)= \sum_{i \in l.o.s.} \frac{N(\lambda,\beta,r,d_{i})}{\Delta \Omega}\frac{\pi r^2}{d_i^2} I_{\gamma}(E_{\gamma},r)
    \label{eq:intensitymodelv0}
\end{equation}
where the summation is extended to all the bodies lying along the line-of-sight (l.o.s.) in the direction $(\lambda,\beta)$. In Eq. \ref{eq:intensitymodelv0} we have indicated with $N(\lambda,\beta,r,d_{i})$ the number of asteroids along the line-of-sight at distance $d_{i}$ from the Earth, with radius $r$.

We can then define the fraction of bodies at distance $d_{i}$ within the cone pointing towards the direction $(\lambda,\beta)$ as:

\begin{equation}
w(\lambda, \beta, r, d_{i}) = \frac{N(\lambda,\beta,r,d_{i})}{N_{tot}(r)}
\label{eq:weight}
\end{equation}
where $N_{tot}(r)$ is the total number of bodies of radius $r$ in the sky. With this definition, Eq.~\ref{eq:intensitymodelv0} can be rewritten as follows:

\begin{multline}
\phi_\gamma(E_{\gamma},r,\lambda,\beta) =  \\
\frac{1}{\Delta \Omega} \sum_{i \in l.o.s.} w(\lambda,\beta,r,d_i)  \frac{\pi r^2}{d_i^2} I_{\gamma}(E_{\gamma},r) N_{tot}(r).
\label{eq:intensitymodelv1}
\end{multline}

If we assume that all SSSBs are equally distributed in the sky, $w$ is independent of $r$. Therefore, Eq.~\ref{eq:intensitymodelv1} can be rewritten as:

\begin{multline}
    \phi_\gamma(E_{\gamma},r,\lambda,\beta) = \\ 
    \pi r^2 I_{\gamma}(E_{\gamma},r) N_{tot}(r)
    \sum_{i \in l.o.s.} \frac{w(\lambda,\beta, d_i)}{\Delta \Omega \, d_i^2} 
    \label{eq:intensitymodel}
\end{multline}

The r.h.s. of Eq.~\ref{eq:intensitymodel} can then be viewed as the product of two factors: a spectral factor, given by the intensity at production for an asteroid of radius $r$, weighted by a factor $\pi r^2$ and by the number of asteroids with radius $r$; and a spatial factor, containing the fraction of bodies at given spatial coordinates, divided by the solid angle $\Delta \Omega$, weighted with the inverse of their squared distance from the Earth and summed along l.o.s. If the spatial factor and $I_{\gamma}(E_{\gamma},r)$ are known or estimated, by fitting the LAT data with the model in Eq.~\ref{eq:intensitymodel}, it is possible to set constraints on the distribution $N_{tot}(r)$.
To build a spatial map of the asteroid emission, we start by estimating the spatial factors for different directions ($\lambda,\beta$). An asteroid orbit is characterized by four parameters:

\begin{itemize}
    \item the orbit major semiaxis $l$;
    \item the inclination angle $i$ of the orbit with respect to the ecliptic plane;
    \item the longitude $\Omega$ of the ascending node (i.e. one of the two intersection points between the asteroid orbit and the ecliptic plane), measured with respect to the direction of the First Point of Aries;
    \item the argument of periapsis $\omega$, i.e. the angle between the ascending node direction and the major semiaxis.
\end{itemize}

\begin{figure}[t]
    \centering
    \includegraphics[width=0.95\columnwidth,clip]{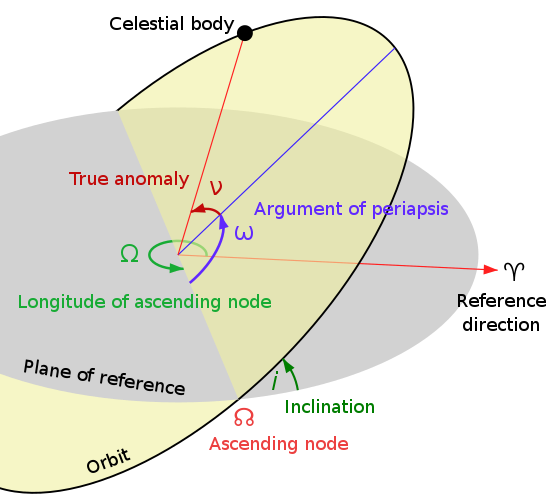}
    \caption{Graphical representation of the orbit of an asteroid~\citep{orbit}. }
    \label{fig:astorbit}
\end{figure}

The asteroid position on its orbit is identified by the angle $\nu$ that its direction forms with respect to the major semiaxis, called ``true-anomaly'' angle. A graphical representation of an asteroid orbit is shown in Figure~\ref{fig:astorbit}.

\begin{figure*}[t]
    \centering
    \includegraphics[width=0.95\columnwidth,clip]{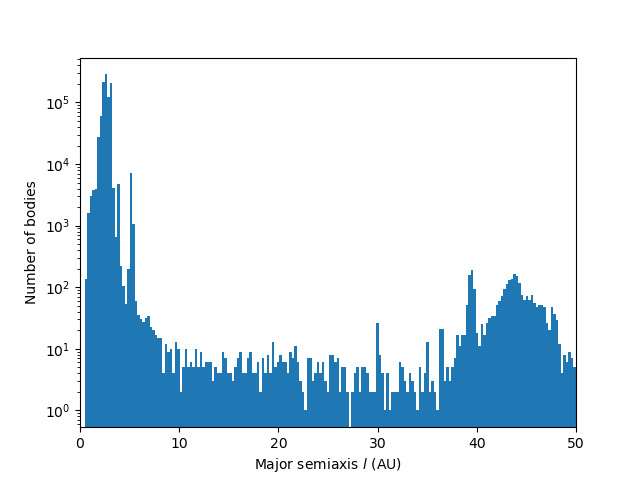}
    \includegraphics[width=0.95\columnwidth,clip]{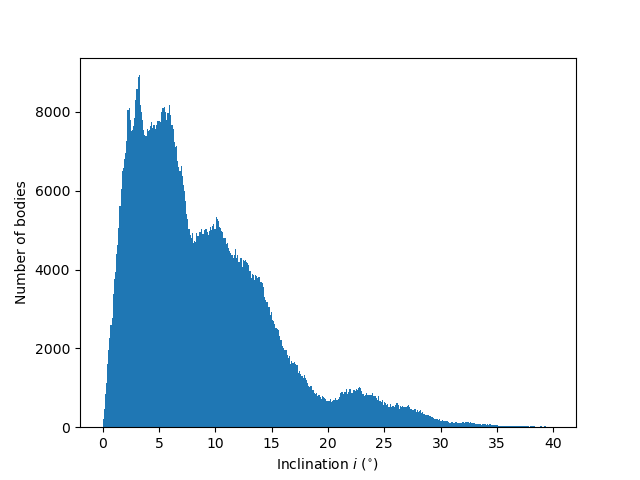}
    \includegraphics[width=0.95\columnwidth,clip]{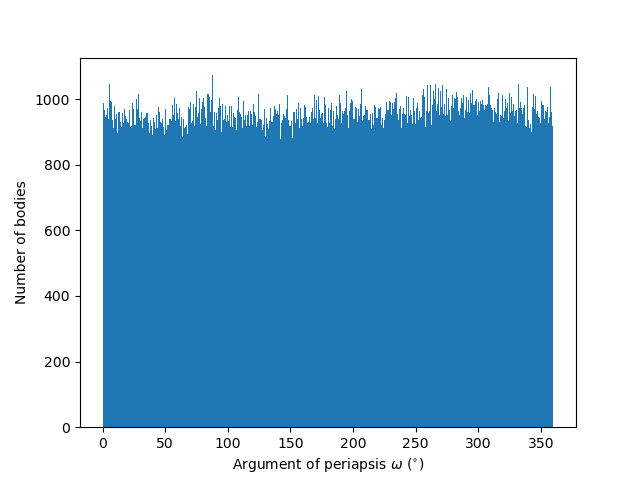}
    \includegraphics[width=0.95\columnwidth,clip]{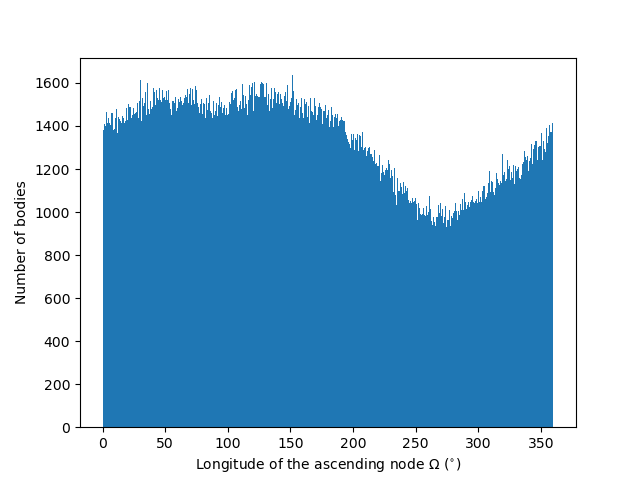}
    \caption{Distributions of the orbital parameters of the SSSBs in the JPL catalog. Top left: major semiaxis; top right: inclination; bottom left: argument of periapsis; bottom right: ascending node longitude.}
    \label{fig:orbitdist}
\end{figure*}

Fig.~\ref{fig:orbitdist} shows the distributions of the orbital parameters of the SSSBs found in the JPL catalog.
We verified that such parameters are not correlated with each other. To build a spatial map of the asteroid population, we sampled $10^{7}$ times the orbital parameters, assuming circular orbits, and extracting the true-anomaly angles from a uniform distribution between $0\degrees$ and $360\degrees$. For each set of parameters $l$, $i$, $\Omega$, $\omega$ and $\nu$, the corresponding asteroid position can be evaluated in a right-handed reference frame centered on the Sun with the $x$-axis directed from the Sun to the First Point of Aries, and the $z$-axis perpendicular to the ecliptic plane. In this frame, the position of the asteroid is given by:

\begin{equation}
    \begin{aligned}
    &x_{A} = l [\cos \Omega \cos(\omega+\nu)- \sin \Omega \cos i \sin(\omega+\nu)] \\
    &y_{A} = l [\sin \Omega \cos(\omega+\nu)- \cos \Omega \cos i \sin(\omega+\nu)] \\
    &z_{A} = l \sin i \sin(\omega+\nu)
    \end{aligned}
    \label{eq:distastsun}
\end{equation}

From Eq.~\ref{eq:distastsun} the ecliptic coordinates of the asteroid can be evaluated:

\begin{equation}
    \begin{aligned}
    &\lambda_{ecl} = \arctan{\frac{y_{A}}{x_{A}}}\\
    &\beta_{ecl} = \arcsin{\frac{z_{A}}{\sqrt{x_{A}^2+y_{A}^2+z_{A}^2}}}
    \end{aligned}
    \label{eq:eclipticcoordinates}
\end{equation}
By following the prescriptions in~\citet{astrocalc}, the ecliptic coordinates of each body can be converted into celestial and galactic coordinates.

If the Sun-Earth direction forms an angle $\xi$ with respect to the x-axis, the Earth coordinates (in $\units{AU}$ units) are given by $x_{E} = \cos{\xi}$, $y_{E} = \sin{\xi}$, $z_{E} = 0$. For each simulated asteroid position we extracted the angle $\xi$ describing the position of the Earth from a uniform distribution between $0\degrees$ and $360\degrees$. The asteroid distance from the Earth is therefore given by $d=\sqrt{(x_{A}-x_{E})^2+(y_{A}-y_{E})^2+(z_{A}-z_{E})^2}$.

The asteroid spatial map is built using a HEALPix\footnote{\url{http://healpix.sourceforge.net}} pixelization of the sky with $N_{side}=32$. This means that the sky is divided into $12 \times N_{side}^2 = 12288$ pixels, each one with solid angle $\Delta \Omega=1.02 \times 10^{-3} \units{sr}$. Each pixel is assigned a weight given by the spatial factor in Eq. \ref{eq:intensitymodel}:

\begin{equation}
    \bar{w}(\lambda,\beta)=\frac{1}{\Delta \Omega}
    \sum_{i \in l.o.s.} \frac{w(\lambda,\beta, d_i)}{d_i^2} 
    \label{eq:astspmap}
\end{equation}
where $\lambda$ and $\beta$ are the coordinates at the center of the given pixel, $N(\lambda,\beta, d_i)$ is the number of objects with distance $d_{i}$ from the Earth whose coordinates are contained in that pixel and $N$ is the total number of simulated asteroids.

\begin{figure}[ht]
    \centering
    \includegraphics[width=0.95\columnwidth,clip]{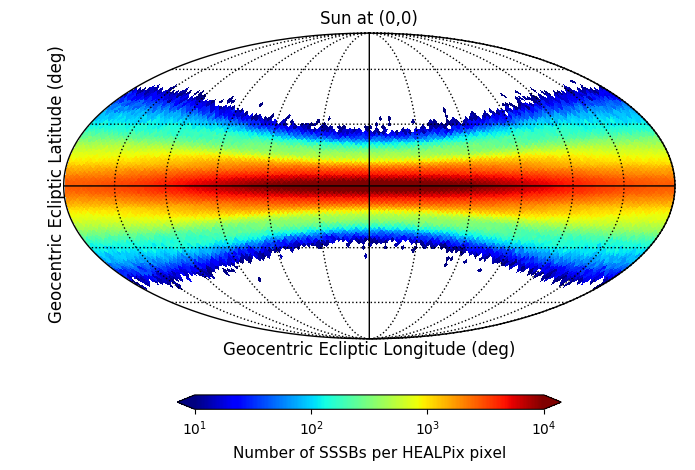}
    \includegraphics[width=0.95\columnwidth,clip]{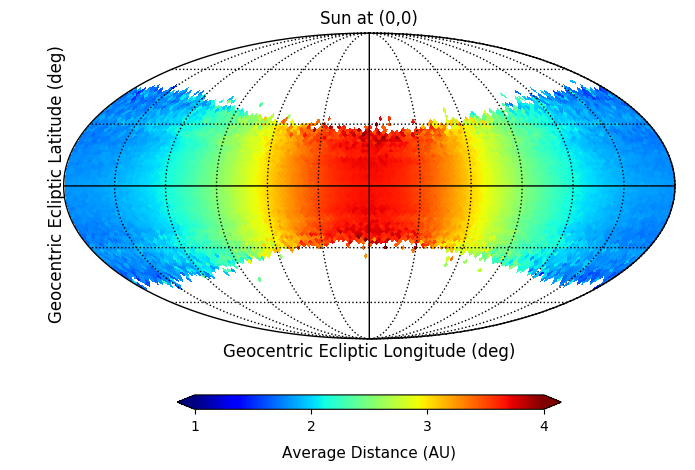}
    \includegraphics[width=0.95\columnwidth,clip]{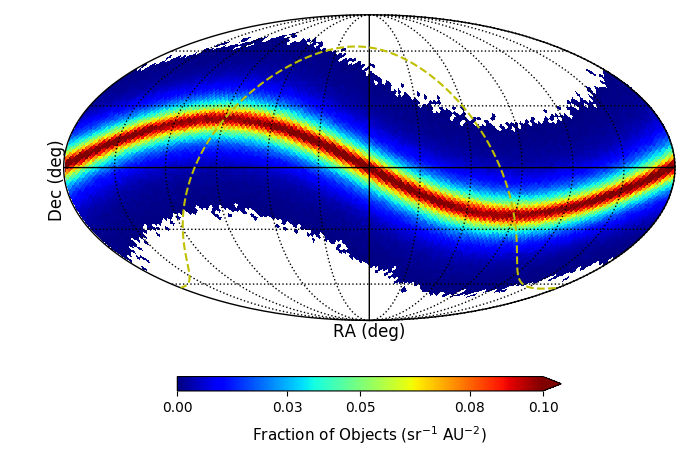}
    \caption{Spatial maps of asteroids obtained from the simulation. Top panel: number of SSSBs per HEALPix pixel in ecliptic coordinates. Middle panel: average asteroid distance from the Earth as a function of its position in the sky. Bottom panel: spatial template of the asteroids (Eq.~\ref{eq:astspmap}) in equatorial coordinates, built using a HEALPix pixelization of the sky with $N_{side}=32$ (Mollweide projection). The dashed yellow line represents the galactic plane.}
    \label{fig:astspmap}
\end{figure}

The asteroid spatial maps are shown in Fig.~\ref{fig:astspmap}.
The top panel shows the asteroid distribution in ecliptic coordinates. The middle panel shows the average asteroid distance from the Earth as a function of the asteroid position. We see that the asteroid distribution is peaked toward the direction of the Sun (which is at the center of the map). This projection effect results from the asteroids along the Earth-Sun direction being, on average, farther from the Earth than those along the opposite direction. Finally, the bottom panel shows the spatial template of Eq.~\ref{eq:astspmap} in celestial coordinates. 

\section{LAT Data Analysis}
\label{sec:analysis}

The LAT is a gamma-ray pair conversion telescope, designed to detect photons in the energy range from $20 \units{MeV}$ up to more than $300 \units{GeV}$. It consists of a $4 \times 4$ array of $16$ identical towers, each one composed of a tracker (TKR) and a calorimeter (CAL) module. Incident gamma rays are converted into $e^+e^-$ pairs, whose energies and directions are measured by the CAL and the TKR, thus providing information on the photon energy and direction. The TKR 
contains 36 alternating layers of silicon strip detectors interleaved with tungsten converter foils, for a total on-axis thickness of $1.5$ radiation lengths. The CAL consists of 96 CsI (Tl) crystals, hodoscopically arranged in 8 layers, for a total on-axis thickness of $8.6$ radiation lengths. The towers are surrounded by a segmented anticoincidence detector (ACD), made of plastic scintillators, working as a veto for charged cosmic rays. Detailed descriptions of the instrument can be found in~\citet{Fermi-LAT:2009ihh} and~\citet{Fermi-LAT:2009gbh,LATtelescope}.

\begin{figure}[!t]
    \centering
    \includegraphics[width=0.95\columnwidth,clip]{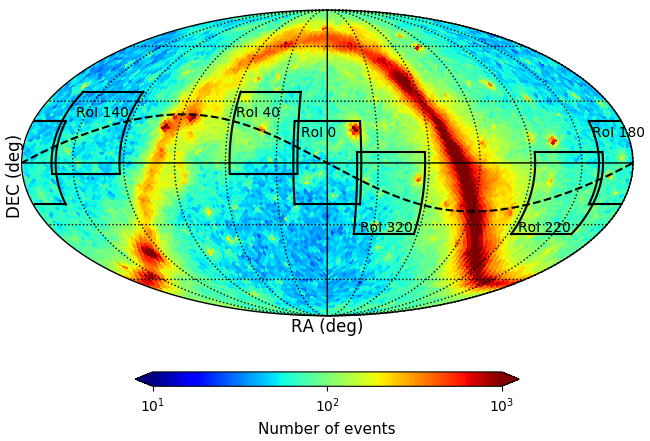}
    \caption{Spatial map in J2000 right ascension and declination of all events detected by the LAT in August 2008 with energy between $56 \units{MeV}$ and $1 \units{TeV}$ and zenith angle $>90\degrees$ (Mollweide projection). The boundaries of the six analysis RoIs are also indicated by the black continuous lines.}
    \label{fig:allskyevt}
\end{figure}

The data sample used for the present analysis has been extracted from the {\tt Pass 8 P305} dataset~\citep{Fermi-LAT:2013jgq}, selecting ULTRACLEANVETO event class (front and back) photons~\footnote{This is the event class with the smallest fraction of residuals (misclassified) cosmic rays and is recommended for studies of diffuse emission (see \url{https://fermi.gsfc.nasa.gov}).}, with energies between $56\units{MeV}$ and $1.78\units{TeV}$, collected in the period from August 2008 (MET=239557418) to December 2020 (MET=631153850)~\footnote{The Mission Elapsed Time, or MET, is the number of seconds since the reference time of January 1, 2001, at 0h:0m:0s in the Coordinated Universal Time (UTC) system, corresponding to a Modified Julian Date (MJD) of 51910 in the UTC system (see \url{https://fermi.gsfc.nasa.gov/ssc/data/analysis/documentation/Cicerone/Cicerone_Data/Time_in_ScienceTools.html}).}. The energy interval has been divided into logarithmic bins, with 8 bins per decade. The analysis has been performed in six different Regions of Interest (RoIs) along the ecliptic plane, of $40\degrees$ width in equatorial latitude and longitude, thus selecting the parts of the sky where the asteroid signal is expected to be maximal. A minimum separation of $\simeq 17\degrees$ from the Galactic Equator was required, in order to avoid the strong contamination from the diffuse interstellar gamma-ray emission in the Milky Way. The resulting RoIs are centered at the ecliptic longitudes $0\degrees$, $40\degrees$, $140\degrees$, $180\degrees$, $220\degrees$ and $320\degrees$, and at the ecliptic latitude $0\degrees$ (see Fig.~\ref{fig:allskyevt}). In the following we will designate these regions as RoI 0, RoI 40, RoI 140, RoI 180, RoI 220 and RoI 320.

We selected the time intervals when the LAT was operating in its standard science operation configuration and was outside the South Atlantic Anomaly (SAA). To limit contamination from the Earth limb, we discarded time intervals when the LAT z-axis was at an angle $>70\degrees$ with respect to the zenith direction. This tight zenith cut has been implemented to take into account the broad instrument point spread function (PSF) below $100 \units{MeV}$. When selecting the good time intervals for the data analysis, we also required a minimum angular separation of $35\degrees$ ($45\degrees$) between the Moon (Sun) direction and the center of each RoI, to avoid contamination from lunar (solar) gamma rays\footnote{We did not implement any cut on the positions of major planets since the LAT has not yet detected emission from any of them.}. Due to the cut on the angular separation from the Sun, each RoI is excluded from coverage for a few months of the year. 

The analysis was performed using the {\tt fermitools} (version 2.0.8)~\footnote{\url{https://fermi.gsfc.nasa.gov/ssc/data/analysis/software/}} and {\tt fermipy} (version 1.0.1)~\citep{Wood:2017yyb} packages. For each of the six RoIs, the analysis was performed separately in each month of each year of the selected data sample. In fact, a possible diffuse signal from the asteroids should be time-dependent, since the relative motion of Earth and asteroids implies variations of the spatial map template of Eq.~\ref{eq:astspmap}, due to changes in the relative distances and in the subtended solid angles. In addition, solar modulation could yield variations observable on yearly/monthly timescales~\citep{Fermi-LAT:2016tkg,DeGaetano:2021IG}.

We have implemented a fitting procedure based on a Poisson maximum likelihood approach. The gamma-ray emission from each RoI is modeled including the standard diffuse background templates developed by the Fermi-LAT collaboration, i.e. the Galactic Interstellar Emission model {\tt gll\_iem\_v07.fits} and the isotropic model~\footnote{\url{https://fermi.gsfc.nasa.gov/ssc/data/access/lat/BackgroundModels.html}}. The point-like and extended sources in each RoI are taken from the fourth catalog of LAT sources 4FGL~\citep{Abdollahi2020}. The normalization parameters of the diffuse models and of all the sources within $25\degrees$ from the center of the RoI and detected with TS$>25$ were fitted. An additional source was added to describe the asteroid emission, as discussed in Section~\ref{sec:asteroidsspectrum}. As previously explained, we used the map in Eq.~\ref{eq:astspmap} and shown in Figure~\ref{fig:astspmap} as a spatial template. To minimize the  assumptions on the spectral shape, a power-law with spectral index 2 was used~\footnote{We assume a positive index for the power-law model since we define it as $dN/dE \propto E^{-\alpha}$.}:

\begin{equation}
    \frac{dN_{\gamma}}{dE} = N_{0,\gamma} \left(\frac{E}{E_0}\right)^{-2}
    \label{eq:powerlaw}
\end{equation}
with $E_0 = 100 \units{MeV}$. Here $N_{0,\gamma}$ is the differential flux at $E=E_0$, in units of $\units{MeV^{-1}cm^{-2}s^{-1}}$ and is the only free parameter in this model.

For each fit, we computed the Test Statistic (TS) for the spatial template representing the diffuse emission due to asteroids, defined as

\begin{equation}
    \units{TS} = -2 \, (\ln \mathcal{L}_{max,0} - \ln \mathcal{L}_{max,1})
    \label{eq:TS}
\end{equation}
where $\mathcal{L}_{max,0}$ is the maximum likelihood value for a model without the source of interest (the ``null hypothesis'') and $\mathcal{L}_{max,1}$ is the maximum likelihood value for a model with the additional source (``alternative hypothesis''), which, in this case, is represented by the asteroids. 
The TS is usually used to estimate the significance of the source. In particular, in the case of a model with one additional degree of freedom with respect to the null hypothesis, the significance is equal to $\sqrt{\rm TS}$.

\begin{figure*}[ht]
    \centering
    \includegraphics[width=0.48\textwidth,clip]{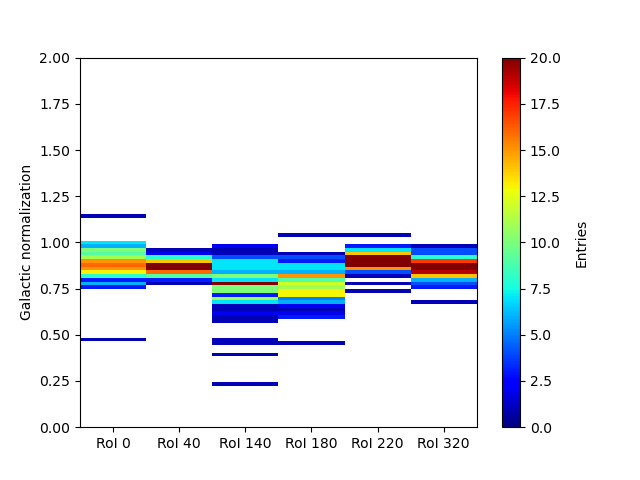}
    \includegraphics[width=0.48\textwidth,clip]{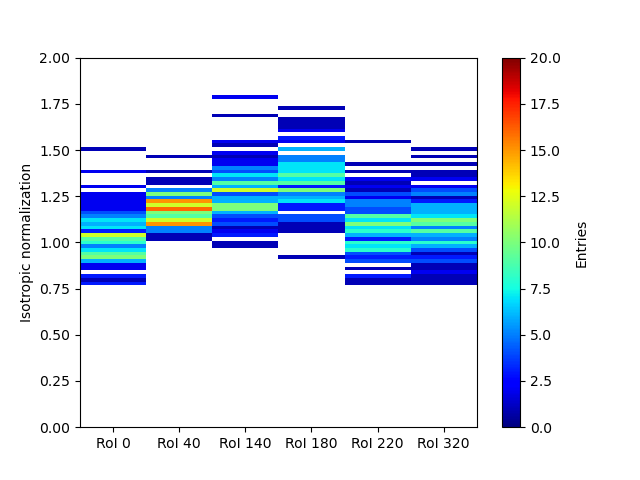}
    \includegraphics[width=0.48\textwidth,clip]{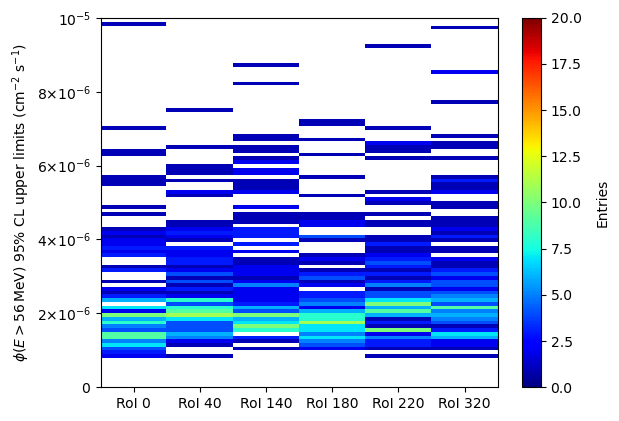}
    \includegraphics[width=0.48\textwidth,clip]{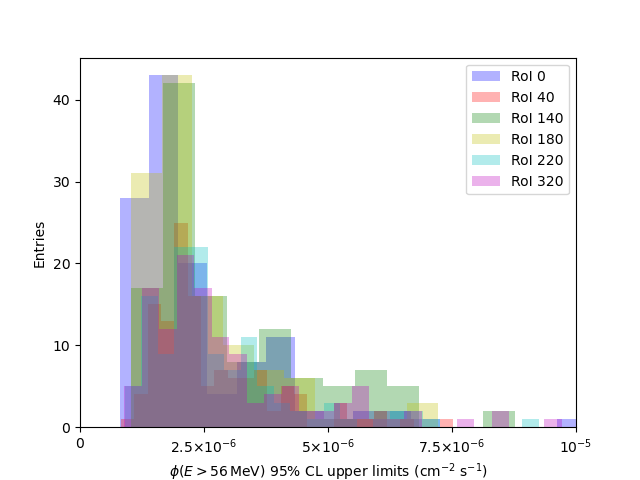}
    \caption{Summary of the fit results. Top left panel: distribution of the values of the normalization constant of the Galactic diffuse background. Top right panel: distribution of the values of the normalization constant of the isotropic background. Bottom left panel: distribution of the ULs at $95\%$ CL on the asteroid gamma-ray flux. Bottom right panel: comparison between the distributions of the ULs at $95\%$ CL in the six RoIs. A total of 741 fits have been performed, corresponding to about 124 fits for each RoI.}
    \label{fig:ulavg}
\end{figure*}

Figure \ref{fig:ulavg} shows a summary of the fit results obtained in the analysis of the different RoIs in the different time intervals. The top panels of Figure \ref{fig:ulavg} show the values of the normalization constants of the Galactic interstellar and of the isotropic diffuse components obtained from the fits. In all fits the TS turned out to be $\simeq 0$, i.e. the asteroid source was not significantly detected for any RoI and time interval. Hence, in each fit we derived the upper limit (UL) on the asteroid flux above $56\units{MeV}$ at $95\%$ confidence level (CL). These limits are shown in the bottom panels of Figure \ref{fig:ulavg}. We see that the distributions of the ULs on the asteroid flux obtained in the various RoIs exhibit similar shapes and are peaked around $2\times 10^{-5} \units{MeV^{-1}cm^{-2}s^{-1}}$. The normalization constants are close to $1$, with the normalization of the Galactic component slightly lower than $1$ and that of the isotropic component slightly exceeding $1$. The two normalization constants also appear to be anticorrelated. 
 
\section{Combined Likelihood analysis}
\label{sec:stackedanalysis}

\begin{figure}[t]
    \centering
    \includegraphics[width=0.95\columnwidth,clip]{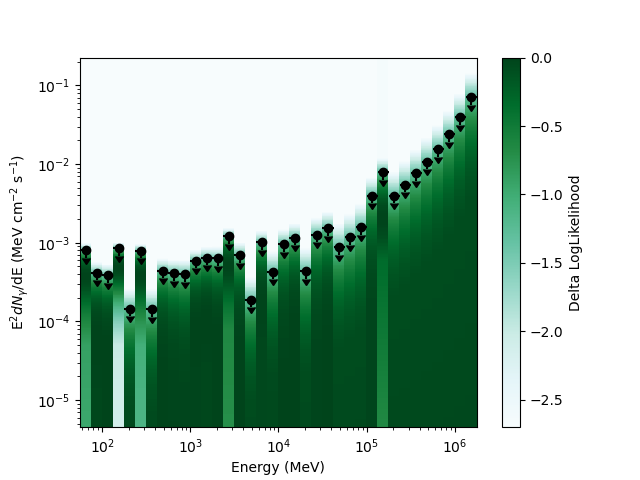}
    \caption{Spectral Energy Distribution (SED) and log-likelihood scan with respect to its maximum value for the asteroids model in RoI 0, August 2008. The color gradient shows the values of $\Delta \ln \mathcal{L}$; the black dots indicate the values of the SED where $\Delta \ln \mathcal{L}=-2.71/2$ and therefore correspond to the ULs at $95\%$ confidence level on the SED.}
    \label{fig:llscan}
\end{figure}

As previously discussed, the analysis was performed for separate RoIs and time intervals over a set of energy bins. In each energy bin we evaluated the likelihood profile for the gamma-ray flux from asteroids using {\tt fermipy}~\footnote{\url{https://fermipy.readthedocs.io/en/latest/advanced/sed.html}}. Figure~\ref{fig:llscan} shows an example of these likelihood profiles for RoI 0 in August 2008; in the figure the values of $\Delta \ln \mathcal{L} = \ln \mathcal{L}_{max} - \ln \mathcal{L}$ are shown as a function of the spectral energy distribution (SED) in the various energy bins from $56\units{MeV}$ to $1.78\units{TeV}$. The likelihood values as a function of the SED are computed by varying the parameter $N_{0,\gamma}$ in Eq.~\ref{eq:powerlaw} and keeping the spectral index fixed to the reference value of $2$ in each energy bin. 

All the likelihood profiles evaluated in the different RoIs, time intervals and individual energy bins have been combined to evaluate constraints on a possible source population with any given spectral shape $f(E)$, with free normalization $C$. Starting from the log-likelihood values $\ln {\mathcal{L}}_i^s$ in the $i$-th energy bin and \mbox{$s$-th} RoI/time bin as a function of the gamma-ray flux $C f(E)$, it is possible to calculate the log-likelihood value $\ln \mathcal{L}_i^s(C)$ for this spectral model in each RoI/time bin. The total log-likelihood for the model is given by:

\begin{equation}
    \ln \mathcal{L}(C) = \sum_s \sum_i \ln \mathcal{L}_i^s(C)
    \label{eq:stackedlikelihood}
\end{equation}

As a starting point, we find the value $C_{max}$ of the normalization constant yielding the maximum likelihood. Then, we evaluate the TS of the model as $-2 \, [\, \ln \mathcal{L}(C=0) - \ln \mathcal{L}(C_{max})\,]$, where $\ln \mathcal{L}(C=0)$ is the log-likelihood value for $C=0$, corresponding to the null hypothesis. The UL at $95\%$ CL on the normalization factor is the value of $C$ for which $\ln \mathcal{L}(C) = \ln \mathcal{L}(C_{max})-2.71/2$.

\begin{figure}[t]
    \centering
    \includegraphics[width=0.95\columnwidth,clip]{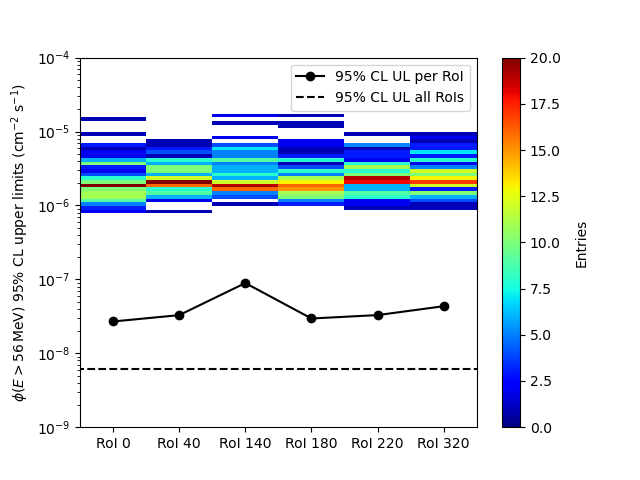}
    \caption{Summary of the ULs at $95\%$ CL on the asteroid flux above $56 \units{MeV}$. The limits obtained from the analysis of individual RoIs in individual one-month time intervals are compared with those obtained from the analysis of individual RoIs combining all time interval (filled circles, continuous line) and with those obtained from the combined analysis of all RoIs and all time intervals (dashed line).}
    \label{fig:fig12}
\end{figure}

This approach is more powerful than the analysis of individual RoIs in restricted time intervals in the search of a possible tiny gamma-ray signal from a population of identical sources. Figure~\ref{fig:fig12} shows the ULs at $95\%$ CL on the power-law fluxes with spectral index $2$ obtained by combining the data for individual RoIs in all time intervals and the data for all RoIs in all time intervals, compared with the limits obtained in the analyses of individual RoIs in individual time intervals. We see that the ULs obtained by combining all the time intervals in an individual RoI are a factor 10 stronger than those obtained in the analysis of the same RoI in an individual time interval. A further improvement of almost a factor 10 is obtained combining the data from all RoIs and all time intervals.

\subsection{Population model-independent analysis: constraints on $N_{tot}(r)$}

We have implemented the analysis procedure illustrated above using for the asteroid source the spectral intensity shape model $f(E)= I_{\gamma}(E,r)$ (see Eq. \ref{eq:intensitymodel}) with fixed values of the asteroid radius. This approach allows for setting constraints on $N_{tot}(r)$ for each value of the radius $r$. We calculated the upper limit on $N_{tot}(r)$ assuming that all asteroids have the same radius and the same composition. 

\begin{figure*}[!ht]
    \centering
    \includegraphics[width=0.95\columnwidth]{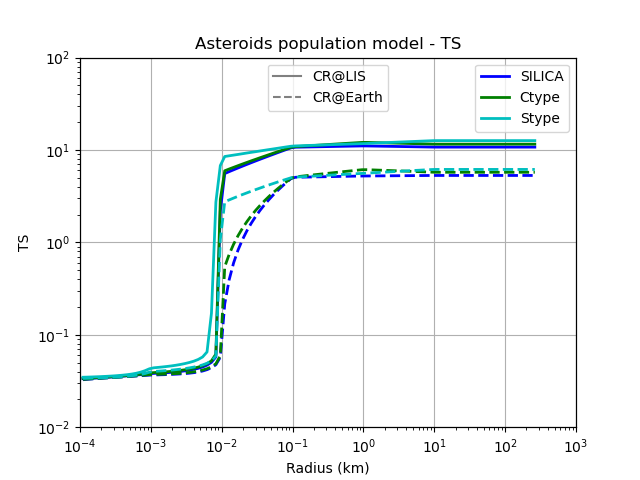}
    \includegraphics[width=0.95\columnwidth]{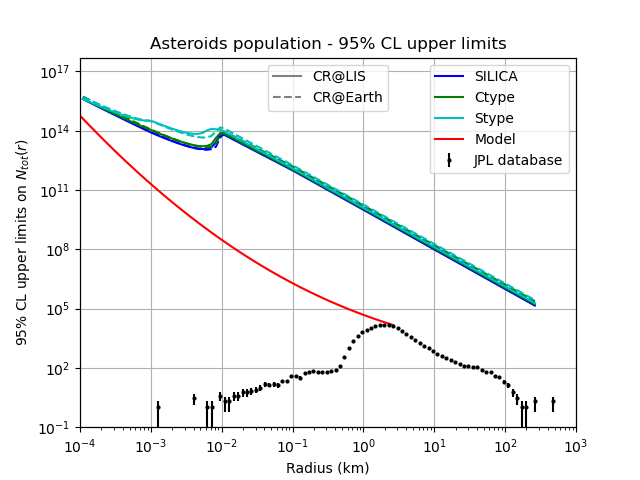}
    \caption{Left panel: TS for the asteroid component as a function of the asteroid radius obtained from the model independent analysis. The values have been calculated for silica, C-type and S-type asteroids, assuming that gamma rays are produced by cosmic rays following either the LIS (blue points) or the spectrum at Earth (green points).
    Right panel: ULs at $95\%$ CL on the total number of asteroids of radius $r$ as a function of the asteroid radius, obtained from the model-independent analysis. 
    The red line represents our population model, which is the extrapolated~\citet{durda1998} model, while the black dots represent the data in the JPL catalog.
    The data and model distributions are binned in radius $r$ with a logarithmic binning of 16 bins per decade.}
    \label{fig:UL_modint}
\end{figure*}

We find that the signal from the asteroids is not significant and we evaluate the constraints on the normalization constant $C$. The UL at $95\%$ CL on the gamma-ray flux, expressed as $\bar{C} I_{\gamma}(E,r)$ can be converted into an upper limit on $N_{tot}(r)$, hereafter indicated as $N_{UL}(r)$, from Eq.~\ref{eq:intensitymodel} integrated over the whole sky:

\begin{equation}
\begin{aligned}
    \phi_{\gamma,\mathrm{UL}}(E,r) &= \bar{C} I_{\gamma} (E,r) = \mathcal{B} \pi r^2 I_{\gamma}(E,r) N_{\mathrm{UL}}(r) \\ 
    \Longrightarrow N_{\mathrm{UL}}(r) &= \frac{\bar{C}}{\mathcal{B} \pi r^2}
\end{aligned}
\label{eq:modelindul}
\end{equation}
where $\mathcal{B}$ is the integral over the entire sky of the asteroid distribution in Eq.~\ref{eq:astspmap} and shown in the bottom panel of Figure~\ref{fig:astspmap}.

In Figure~\ref{fig:UL_modint} we show the TS and the ULs at $95\%$ CL on the total number of asteroids obtained in the hypothesis that all bodies in the population have the same radius $r$, assuming for $I_{\gamma}(E,r)$ the spectra produced by cosmic rays following either the LIS or the spectrum at Earth detected by the AMS-02 experiment interacting with silica, C-type or S-type bodies. The dependence of the TS and of the ULs on $r$ is determined  by the shape of the function $I_{\gamma}(E,r)$. For $r > 10\units{m}$, the TS is almost constant, since the gamma-ray intensity becomes independent of the asteroid radius and $N_{\mathrm{UL}}(r)$ scales as $r^{-2}$. This also explains the slight increase in the ULs for $r\sim 10 \units{m}$. For smaller radii, the spectral shape of the gamma-ray intensity is harder, and it is more disfavored by the data, resulting in a TS closer to zero. The constraints obtained with the different classes of asteroids are similar for radii below $10 \units{m}$ and above $100\units{m}$.
Since the TS does not exceed the value of $\sim 10$, the asteroid source is not significantly detected for any value of $r$ tested in the present work. In the right panel of Figure~\ref{fig:UL_modint}, we also show the number of SSSBs in the JPL catalog and the one predicted by our model (see Section \ref{sec:smallsolarsystembodies}). The distribution is calculated using a logarithmic binning in radius with 16 bins per decade. We remark that the UL obtained with this procedure cannot be directly compared with the assumed size distribution, since each upper limit is derived in the hypothesis that all bodies have the same radius $r$. 

\subsection{Model-dependent analysis: constraints on $N_{tot}(r\geq r_{min})$}

We also applied the combined likelihood analysis technique to set a constraint on the asteroids population model proposed in Section~\ref{sec:smallsolarsystembodies} as an extrapolation of the model of~\cite{durda1998} for $r\leq1.25 \units{km}$ and following the JPL catalog distribution for $r\geq1.25 \units{km}$. 

The cumulative flux of asteroids can be evaluated from Eq. ~\ref{eq:intensitymodel} by summing over all the radii from $r_{min}$ up to $r_f \simeq 300 \units{km}$:

\begin{equation}
    \phi_{\gamma,c} (E) = \mathcal{B} \sum_{r=r_{min}}^{r_f} \pi r^2 I_{\gamma}(E,r)N_{tot}(r).
    \label{eq:cumulativeintensitymodel}
\end{equation}

For a given value of $r_{min}$, the function $f(E)$ is then given by Eq.~\ref{eq:cumulativeintensitymodel}. The ULs on the normalization factor $C$ are then converted into ULs on the cumulative population model. 

\begin{figure*}[!ht]
    \centering
    \includegraphics[width=0.95\columnwidth]{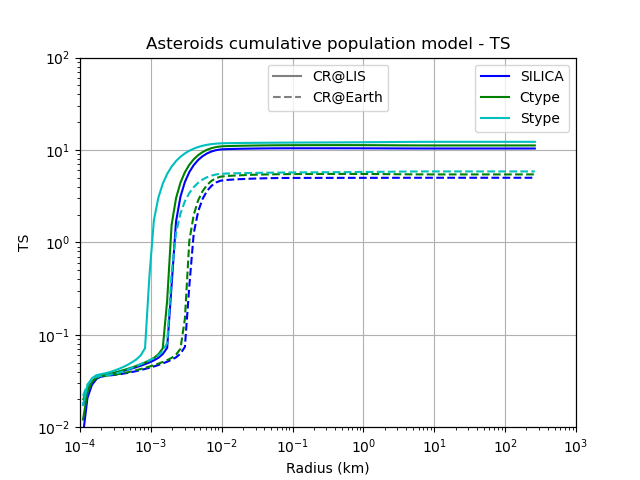}
    \includegraphics[width=0.95\columnwidth]{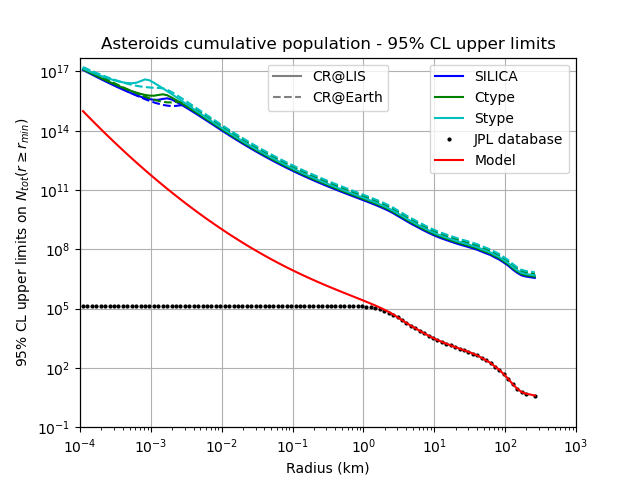}
    \caption{Left panel: TS for the asteroid component as a function of the minimum asteroid radius obtained from the model dependent analysis. The values have been calculated for silica, C-type ans S-type asteroids, assuming that gamma rays are produced by cosmic rays following either the LIS (blue points) or the spectrum at Earth (green points).
    Right panel: ULs at $95\%$ CL on the asteroids population as a function of the  asteroid radius obtained from the model dependent analysis. 
    The red line represents our population model, which is the extrapolated ~\citet{durda1998} model, while the black dots represent the data in the JPL catalog.
    The data and model distributions are binned in radius $r$ with a logarithmic binning of 16 bins per decade.}
    \label{fig:UL_moddip}
\end{figure*}

The left panel of Fig.~\ref{fig:UL_moddip} shows the TS of the model as a function of the minimum radius $r_{min}$ for silica, C-type and S-type asteroids, evaluated assuming either the LIS or the spectra of cosmic rays at Earth measured by AMS-02. The TS  is approximately null for $r_{min}<10^{-3} \units{km}$, while it increases for larger values of $r_{min}$, due to the change of shape of the gamma-ray intensity at the production. Compared with the results in the previous section, the increase is smoother, since, for each value of $r_{min}$, $f(E)$ is obtained from a folding of all the energy spectra of asteroids with $r\geq r_{min}$. At some point the TS reaches a limiting value $\sim 10$, still not significant. As already stated in the previous section, this behavior is due to the fact that for large radii the spectral shape becomes independent of the asteroid size.

In the right panel of Fig. \ref{fig:UL_moddip}, the ULs at $95\%$ CL for the integral population of asteroids with $r \geq r_{min}$ are shown. The limits are above the model in the whole range of $r_{min}$ and the ratio between the UL and the population predicted by the model increases with $r_{min}$, from $\approx 10^2$ for $r_{min}=r_0=10\units{cm}$ to $\approx 10^6$ for $r_{min}=r_{f}\approx 300 \units{km}$.

\begin{figure*}[!th]
    \centering
    \includegraphics[width=0.95\columnwidth,height=0.22\textheight,clip]{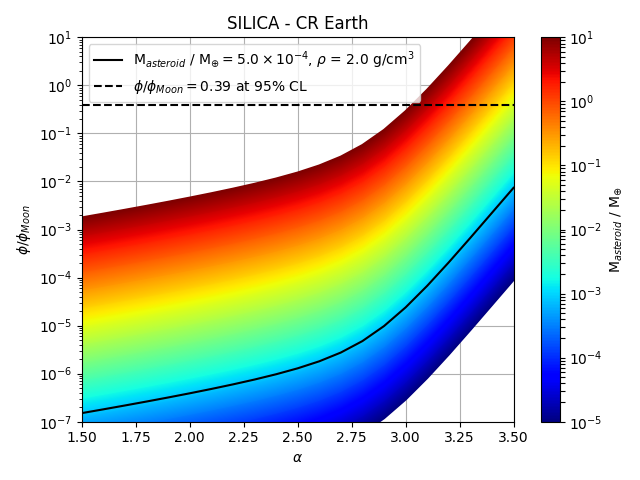}
    \includegraphics[width=0.95\columnwidth,height=0.22\textheight,clip]{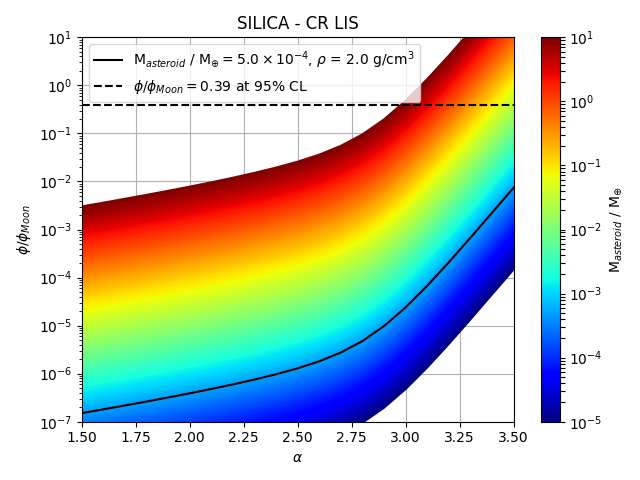}
    \caption{Ratio between the asteroid flux and the Moon flux calculated with Eq. \ref{eq:astcumflux}, as a function of the asteroid size distribution power-law index $\alpha$, for different values of the asteroid total mass. All asteroids have been assumed to lie at a distance $d=2.7 \units{AU}$ from the Earth. The gamma-ray intensity at production is obtained simulating the interaction of CRs with  silica bodies. The plot in the left panel is obtained using the CR spectra measured at Earth, while the plot in the right panel is obtained using the CR LIS. The continuous black lines indicate the results obtained assuming for the total asteroid mass the value $5 \times 10^{-4} M_{\oplus}$. The dashed lines correspond to the measured UL at $95\%$ CL. }
    \label{fig:ul_pl_CR}
\end{figure*}

\subsection{Comparison between asteroids and Moon flux}
\label{subsec:stackedanalysisresults}

We repeated the combined likelihood analysis using for the spectral model $f(E)$ the Moon gamma-ray flux measured by the LAT in its first seven years of operation~\citep{Fermi-LAT:2016tkg}. The UL at $95\%$ CL on the asteroid flux $\phi_{A}^{UL}$ is 0.39 times the Moon flux.

We have evaluated the ratio between the asteroid flux and the Moon flux $\phi/\phi_{Moon}$ using Eq. \ref{eq:astcumflux} under the following assumptions: (i) the asteroids are all located at a distance $d=2.7\units{AU}$ from the Earth; (ii) the asteroids are all composed of silica with a density of $2\units{g/cm^{3}}$; (iii) the asteroid size distribution follows a power-law with index $\alpha$, according to Eq. \ref{eq:dndr}. The integration limits were fixed to the values $r_0 = 1 \times 10^{-4} \units{km}$ and $r_1 = 470  \units{km}$. Figure ~\ref{fig:ul_pl_CR} shows the ratio $\phi/\phi_{Moon}$ as a function of the power-law index $\alpha$ for different values of the total asteroid mass. In the left panel the gamma-ray flux from asteroids is evaluated using the CR spectra measured at Earth, while in the right panel it is evaluated using the CR LIS. The dotted line represents the value obtained from the combined analysis performed using the average Moon flux as spectral model. At $95\%$ CL, all values of $\alpha$ and asteroid total masses above this threshold can be ruled out.

\begin{figure}[!th]
    \includegraphics[width=0.95\columnwidth,height=0.22\textheight,clip]{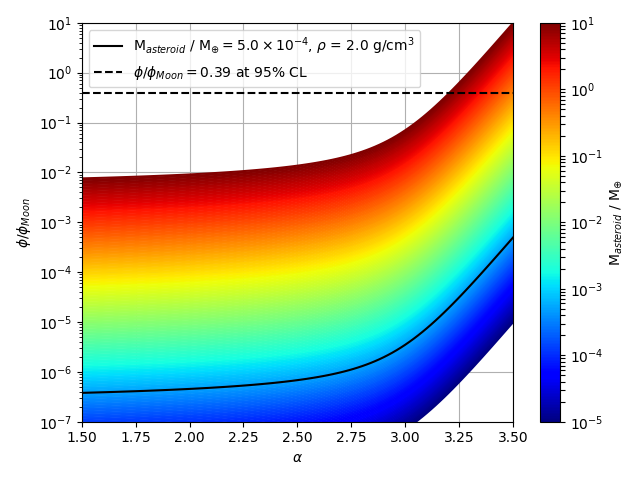}
    \caption{Ratio between the asteroids flux and the Moon flux calculated with Eq. \ref{eq:astflux1} as a function of the asteroid size distribution power-law index $\alpha$, for different values of the asteroid total mass.}
    \label{fig:ul_pl}
\end{figure}

In Figure \ref{fig:ul_pl}, the same ratio is shown as calculated in Eq.  \ref{eq:ratioastmoonflux}, i.e. assuming the gamma-ray intensity from asteroids at production to be equal to the gamma-ray intensity from the Moon at production. When using Eq. \ref{eq:ratioastmoonflux}, the ratio $\phi/\phi_{Moon}$ is independent of energy, and depends only on geometrical parameters. 

\section{Conclusions}

In this work we have used the LAT Pass 8 ULTRACLEANVETO data collected from August 2008 to December 2020 to study the gamma rays from the ecliptic plane with energies in the range $56 \units{MeV} - 1 \units{TeV}$, aiming to constrain the gamma-ray emission from SSSBs. Such bodies are mostly located around the ecliptic plane and are expected to act as passive sources of gamma rays due to the interactions of cosmic rays with their surfaces. This method provides a unique way to constrain the population of all small bodies with diameters $< 2 \units{m}$, on which no data are available in the JPL catalog of SSSBs~\citep{JPL}. 

In addition, this analysis allows for testing different models of the population of asteroids. In particular, we have used the analysis results to constrain a size distribution model (in the hypothesis that asteroids are spherical) that we have built by extrapolating the \citep{durda1998} model for diameters down to $\simeq 20 \units{cm}$ and using the JPL catalog for diameters larger than $2.5\units{km}$. 

The analysis was performed in six different RoIs at a distance $>17^{\circ}$ from the Galactic plane to avoid the bright contamination of the latter. Separate analyses were performed for individual RoIs in each year and each month of the selected sample. This choice was motivated to search, in case of a source detection, for a signal modulated by solar activity over 12 years of data, and by the changes of the distance between the Earth and the asteroids as the Earth orbits the Sun.  

The gamma-ray emission from each RoI was modeled including the standard diffuse background templates developed by the Fermi-LAT collaboration and the point-like and extended sources from the 4FGL catalog. An additional source describing the asteroids diffuse emission was implemented. Its morphology was built by sampling $10^7$ times the asteroids orbital parameters from the JPL SSSBs catalog, while its spectral emission was modeled as a power-law of index 2. The analysis was performed with free normalizations of the most significant sources in each RoI and the prefactor $N_0$ of the asteroids power-law energy spectrum. For each fit, we computed the TS of the source and we found that the asteroids source was not detected.

As a second step, we computed the SED of the asteroids in each RoI-time bin in which the data sample was divided, and we implemented a combined likelihood analysis to constrain the asteroids population. We assumed the asteroids flux to be given by the model in Eq.~\ref{eq:intensitymodel}, i.e. by the convolution of the asteroids spatial map and their intensity at production level, weighted by the number $N(r)$ of asteroids with radius $r$ and the factor $\pi r^2$. The intensity at the production site was computed with the {\tt Fluka} code by simulating the interactions of charged cosmic rays with bodies of different radius. As for the CR spectra, we adopted both a spectrum near the Earth, measured by the AMS-02 experiment, and the LIS spectrum, taken from~\citet{DeLaTorreLuque:2021yfq,Luque:2021nxb,Luque:2022aio}, and we assumed 
different asteroid composition models. We tested both homogeneous compositions and more realistic ones representing the most abundant species of asteroids.
The TS of the possible asteroid source is $\lesssim 10$ for any model considered, which corresponds to a significance of approximately $3 \sigma$, insufficient for claiming a detection. We used the model of Eq.~\ref{eq:intensitymodel} to convert the ULs on the flux into ULs on  $N_{tot}(r)$, assuming that all asteroids have the same radius and composition. We found that the population ULs at $95\%$ CL vary between $4.0 \times 10^{15}$ for $r \simeq 10^{-4} \units{km}$ and $1.4 \times 10^5$ for $r \simeq 300 \units{km}$ with the LIS spectrum, and between $5.2 \times 10^{15}$ for $r \simeq 10^{-4} \units{km}$ and $2.0 \times 10^5$ for $r \simeq 300 \units{km}$) with the Earth spectrum.

Then, we used the combined likelihood analysis to constrain the cumulative population of asteroids in our model, assuming the asteroids flux given by Eq.~\ref{eq:cumulativeintensitymodel}. Again, we computed the TS of this model and we found again 10.5, which is still not significant. The ULs at $95\%$ CL on the cumulative population of asteroids are about $100$ times larger than the predictions of the model for $r \simeq 10^{-4} \units{km}$. 

We remark here that the simulation code can be customized to model any asteroid composition and density. Nonetheless, the present LAT data analysis showed that the constraints on the asteroid population do not significantly change when using different asteroid composition models, mainly due to the current LAT sensitivity for this gamma-ray extended source.

Finally, we repeated the combined likelihood analysis by assuming the asteroids flux to be given by the average Moon flux measured by the LAT in its first seven years of operation. In the hypothesis that the asteroids size distribution is described by a power-law of index $\alpha$, the UL of the flux provides a threshold to the values that the asteroids mass and $\alpha$ can assume (see Eqs. \ref{eq:a}, \ref{eq:astflux1} and \ref{eq:ratioastmoonflux}). This comparison is motivated by the fact that the gamma-ray emission of individual asteroids is expected to be similar to that of the Moon, once the proper differences in terms of composition, density and size are taken into account. We found an upper limit at $95\%$ CL of 0.39 for the ratio between the asteroids and the Moon fluxes. Assuming that all asteroids are composed of silica with a density of $2 \units{g/cm^3}$ and are at a distance from the Earth of $2.7 \units{AU}$, this constrains the asteroids mass and $\alpha$ to assume all values below the dashed line in Figures \ref{fig:ul_pl_CR} and \ref{fig:ul_pl}.

\section*{Acknowledgments}
The Fermi LAT Collaboration acknowledges generous ongoing support from a number of agencies and institutes that have supported both the development and the operation of the LAT as well as scientific data analysis. These include the National Aeronautics and Space Administration and the Department of Energy in the United States, the Commissariat \`a l'Energie Atomique and the Centre National de la Recherche Scientifique / Institut National de Physique Nucl\'eaire et de Physique des Particules in France, the Agenzia Spaziale Italiana  and the Istituto Nazionale di Fisica Nucleare in Italy, the Ministry of Education, Culture, Sports, Science and Technology (MEXT), High Energy Accelerator Research Organization (KEK) and Japan Aerospace Exploration Agency (JAXA) in Japan, and the K.~A.~Wallenberg Foundation, the Swedish Research Council and the Swedish National Space Board in Sweden.
 
Additional support for science analysis during the operations phase is gratefully acknowledged from the Istituto Nazionale di Astrofisica in Italy and the Centre National d'\'Etudes Spatiales in France. This work performed in part under DOE Contract DE-AC02-76SF00515.


\vspace{4mm}
\facility{\it Fermi (Fermi-LAT)}

\software{python: \url{https://www.python.org/}, \cite{10.5555/1593511}; matplotlib: \url{https://matplotlib.org/}, \cite{Hunter:2007}; ROOT: \url{https://root.cern/}, \cite{ROOT}; HEALPix, healpy: \url{http://healpix.jpl.nasa.gov/}, \url{https://healpix.sourceforge.io/}, \citep{Gorski:2004by}; FLUKA: \url{http://www.fluka.org/fluka.php}, \cite{Ferrari:2005zk,BOHLEN2014211,BATTISTONI201510}; DPMJET: \cite{10.1007/978-3-642-18211-2_166}; PEANUT: \cite{Fasso:2000hd,battistoni2006recent}}.

\vspace{8mm}

\appendix

\section{Yield and Intensity at production}
\label{sec:appendix}
\begin{figure}
    \centering
    \includegraphics[width=0.44\columnwidth,height=0.67\textheight]{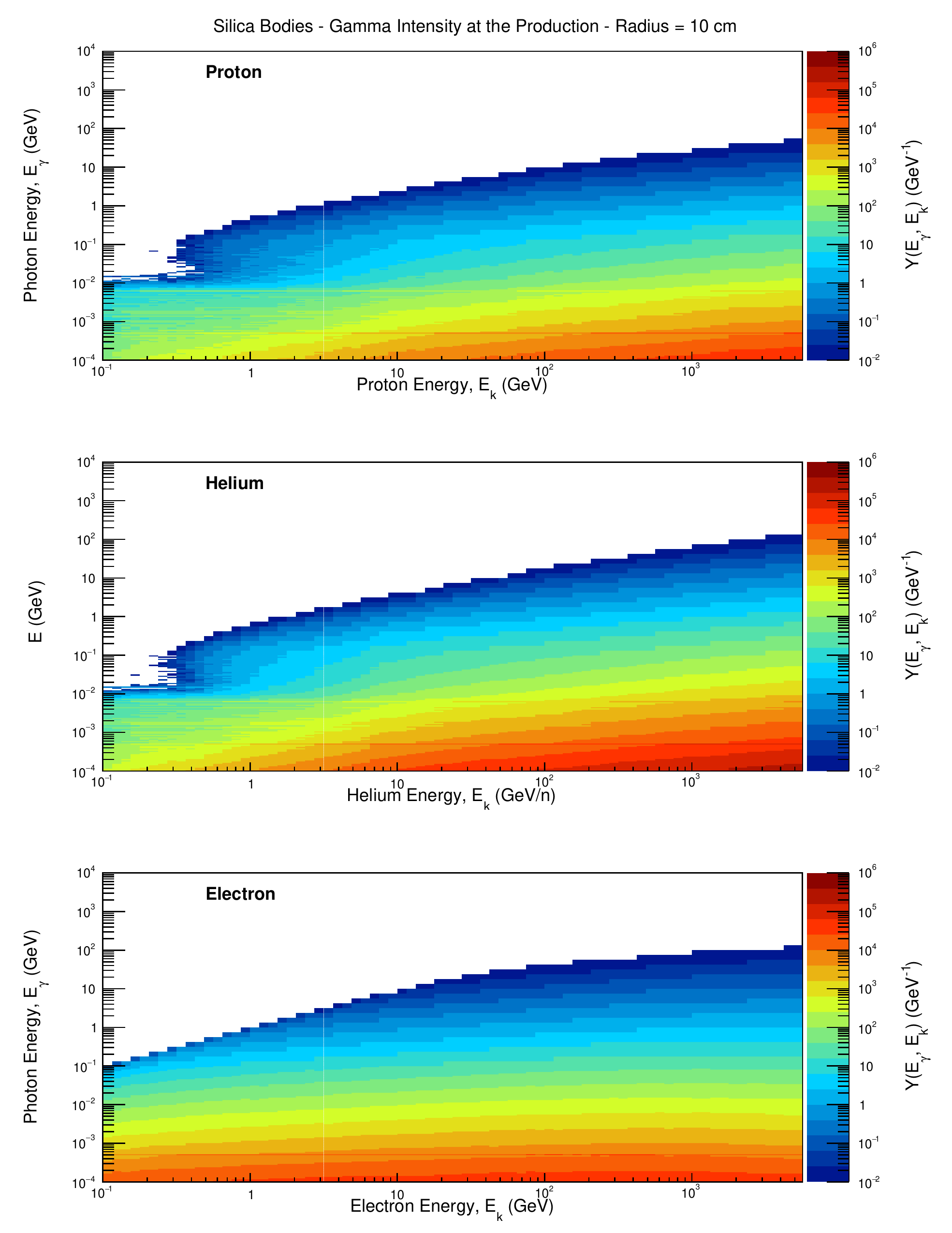}
    \includegraphics[width=0.44\columnwidth,height=0.67\textheight,clip]{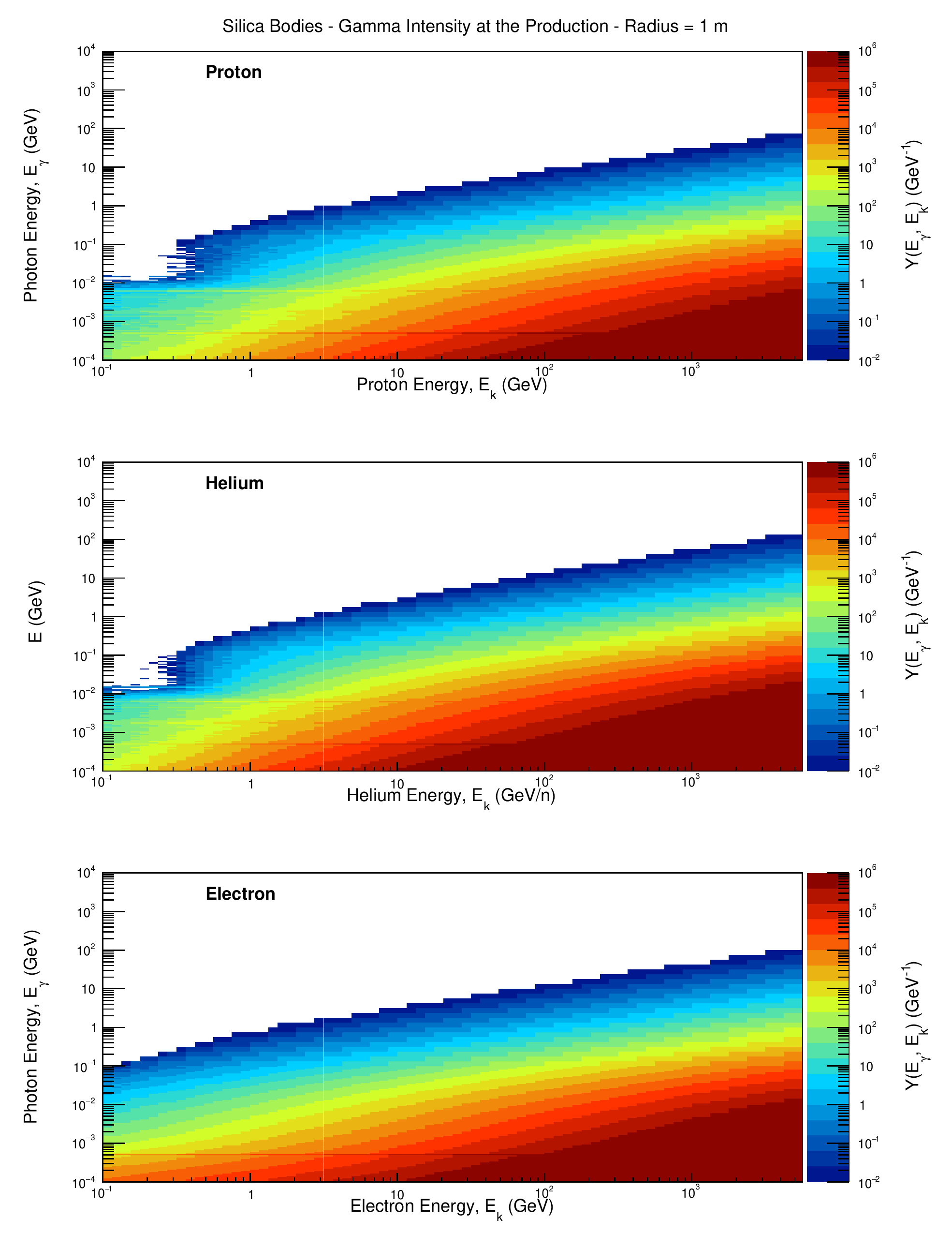}
    \includegraphics[width=0.44\columnwidth,height=0.22\textheight]{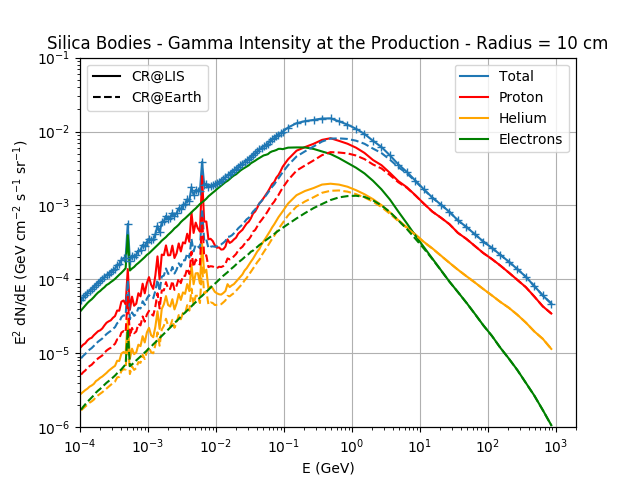}
    \includegraphics[width=0.44\columnwidth,height=0.22\textheight,clip]{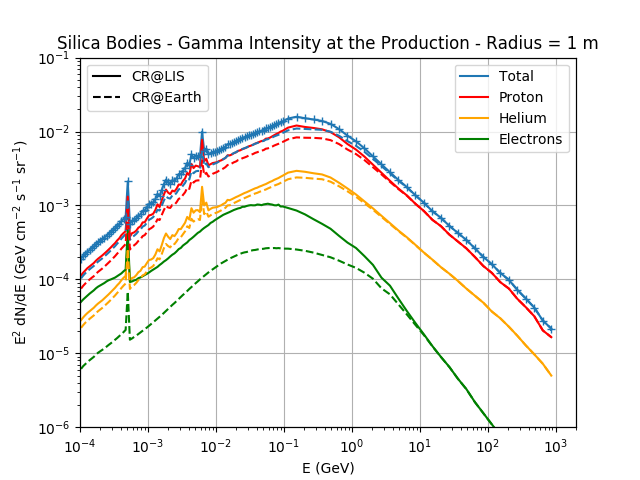}
    \caption{Yields of gamma rays produced by the interactions of protons, $^4$He and electrons with a silica body. The bottom plots show the gamma-ray intensities at the production obtained using the CR LIS spectra (continuous lines) and the CR spectra measured at Earth (dashed lines). Left column: body radius of $10\units{cm}$; right column: body radius of $1\units{m}$.}
    \label{fig:yield_10cm_1m}
\end{figure}

\begin{figure}
    \centering
    \includegraphics[width=0.44\columnwidth,height=0.67\textheight]{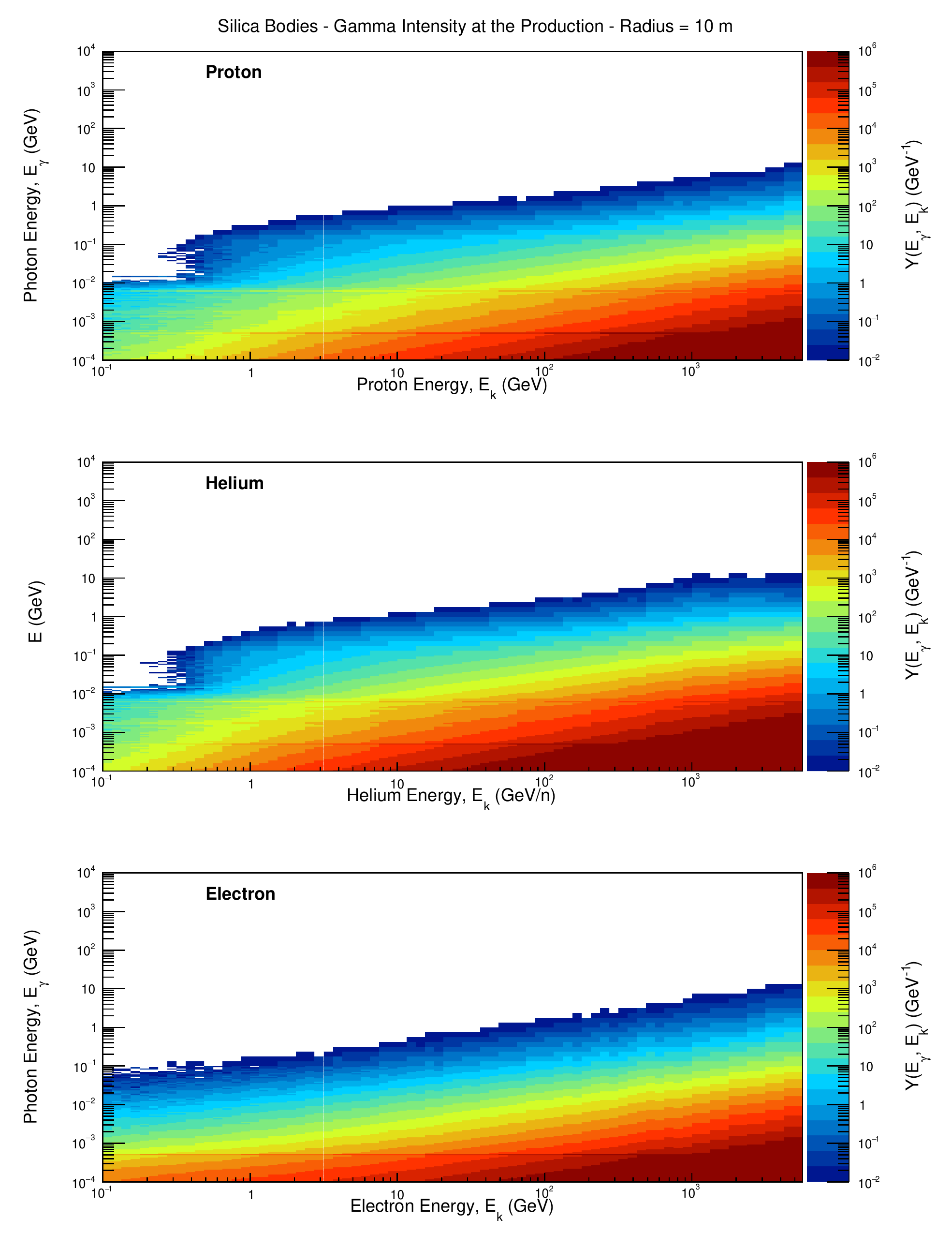}
    \includegraphics[width=0.44\columnwidth,height=0.67\textheight,clip]{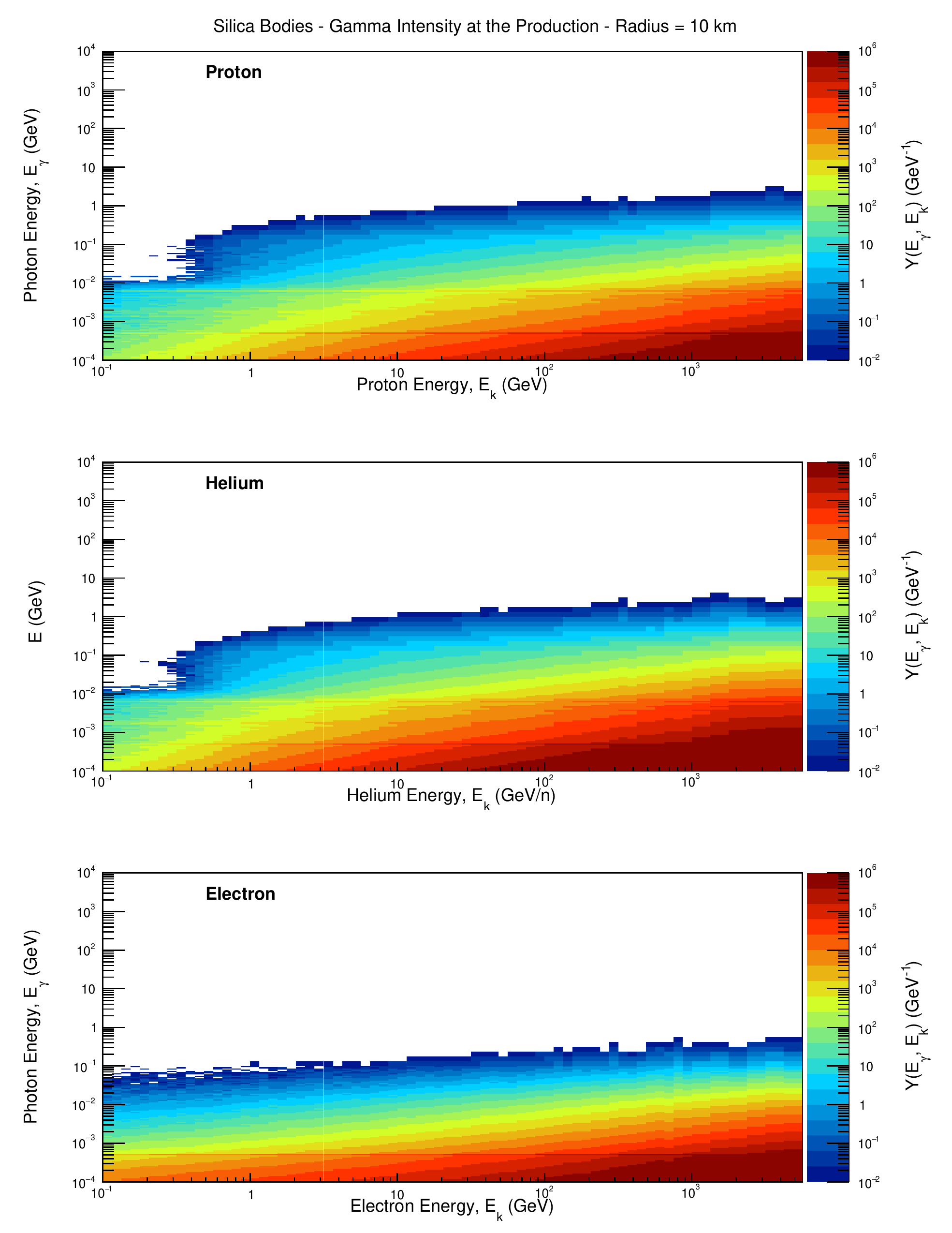}
    \includegraphics[width=0.44\columnwidth,height=0.22\textheight]{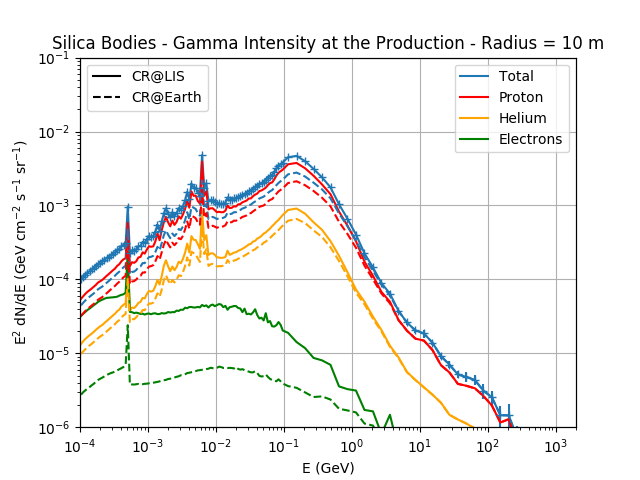}
    \includegraphics[width=0.44\columnwidth,height=0.22\textheight,clip]{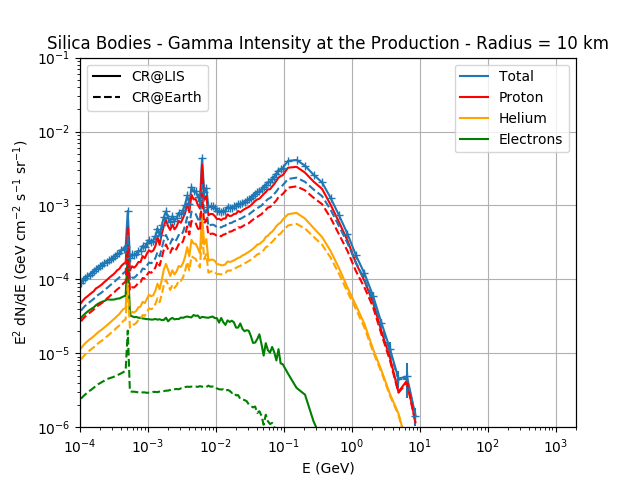}
    \caption{Yields of gamma rays produced by the interactions of protons, $^4$He and electrons with a silica body. The bottom plots show the gamma-ray intensities at the production obtained using the CR LIS spectra (continuous lines) and the CR spectra measured at Earth (dashed lines). Left column: body radius of $10\units{m}$; right column: body radius of $10\units{km}$.}
    \label{fig:yield_10m_10km}
\end{figure}

Figure \ref{fig:yield_10cm_1m} and Figure \ref{fig:yield_10m_10km} show the gamma-ray yields calculated with \texttt{FLUKA}, produced in the interactions of cosmic-ray protons, helium nuclei and electrons with silica bodies of different radii, from $10\units{cm}$ to $10\units{km}$. The yields have been calculated on a grid of primary energies from $100\units{MeV/n}$ up to $10\units{TeV/n}$ with a spacing of 16 bins per decade and of gamma-ray energies from $0.1\units{MeV}$ up to $100\units{MeV}$ with a spacing of 32 bins per decade and from $100 \units{MeV}$ up to $10\units{TeV}$ with a spacing of 8 bins per decade. 

The bottom panels of the figures show the corresponding gamma-ray intensities at production sites, evaluated by folding the gamma-ray yields with the spectra of the cosmic-ray species interacting with the asteroids.
The contributions to the gamma-ray intensities from individual cosmic-ray species are also shown. We have performed this calculation assuming for different CR species the energy spectra measured at Earth or the Local Interstellar Spectra (LIS). The latter have been taken from~\citet{DeLaTorreLuque:2021yfq,Luque:2021nxb,Luque:2022aio}, while for the spectra at Earth we used the AMS02 measurements (see the text for more details). We see that the average energy of gamma rays produced by each cosmic-ray species decreases as the asteroid radius increases. This feature becomes relevant for radii $>1\units{m}$; correspondingly, the gamma-ray intensities at production from each species become softer. Finally, the error bars (shown only for the total intensities) represent the statistic uncertainties due to the finite number of CR events used in the simulation to evaluate the yields.

\bibliography{references}
\end{document}